\pdfoutput=1
\documentclass[journal]{IEEEtran}
\renewcommand{\vec}[1]{\ensuremath{\boldsymbol{#1}}} 
\usepackage[german,english]{babel}
\usepackage[latin1]{inputenc}
\usepackage{graphicx}
\usepackage{eurosym}
\usepackage[process=auto,cleanup={.dvi,.ps,.pdf,.log}]{pstool}
\usepackage{acronym}
\usepackage{amsfonts}

\makeatletter
\let\MYcaption\@makecaption
\makeatother

\usepackage[font=footnotesize]{subcaption}

\makeatletter
\let\@makecaption\MYcaption
\makeatother

\usepackage{algorithm}
\usepackage{algpseudocode}
\algdef{SE}[DOWHILE]{Do}{doWhile}{\algorithmicdo}[1]{\algorithmicwhile\ #1}%
\makeatletter
\let\OldStatex\Statex
\renewcommand{\Statex}[1][3]{%
	\setlength\@tempdima{\algorithmicindent}%
	\OldStatex\hskip\dimexpr#1\@tempdima\relax}
\makeatother

\usepackage[cmex10]{amsmath}
\usepackage{relsize}
\usepackage{tabularx}
\newcommand{\PreserveBackslash}[1]{\let\temp=\\#1\let\\=\temp}
\let\PBS=\PreserveBackslash
\renewcommand\IEEEkeywordsname{Keywords}
\begingroup
\makeatletter
\catcode`\#=11
\gdef\pstool@bitmap@opts{%
	-dAutoFilterColorImages#false
	-dAutoFilterGrayImages#false %
	-dColorImageFilter#/FlateEncode %
	-dGrayImageFilter#/FlateEncode 
}
\gdef\pstool@pspdf@opts{%
	-dPDFSETTINGS#/prepress %
	-dCompatibilityLevel#1.3 %
	-dEmbedAllFonts#true %
	-dSubsetFonts#true
}
\endgroup
\graphicspath{{figures/}}
\PassOptionsToPackage{hyphens}{url}\usepackage[hyperindex,
linktocpage,
bookmarksnumbered,
plainpages=false,
pdfpagelabels,
colorlinks,
bookmarks,
linkcolor=black,
urlcolor=black,
citecolor=black,
pdfpagemode=UseOutlines,
hyperfootnotes = false,
breaklinks=true]{hyperref}

\ifCLASSINFOpdf
\else
\fi
\usepackage{array}
\usepackage{fixltx2e}

\usepackage{cite}

\begin{document}
%
\title{Optimal Placement and Sizing of Distributed Battery Storage in Low Voltage Grids using Receding Horizon Control Strategies}



\author{ \IEEEauthorblockA{Philipp Fortenbacher, Andreas Ulbig, and  G\"oran Andersson
	}
	\\ 
		\thanks{This publication is an outcome of the research project \textit{Smart Planning} (SFOE~R\&D~contract SI/501190-01). The authors would like to thank the Swiss Federal Office of Energy (SFOE) for the project financing.
			
			P. Fortenbacher, A. Ulbig, and G. Andersson  are with the Power Systems Laboratory, ETH Zurich, Switzerland (e-mail: \{fortenbacher, ulbig, andersson\}@eeh.ee.ethz.ch). 	
}}


\acrodef{LV}[LV]{Low Voltage}
\acrodef{AC-OPF}[AC-OPF]{AC Optimal Power Flow}
\acrodef{OPF}[OPF]{Optimal Power Flow}
\acrodef{FBS-OPF}[FBS-OPF]{Forward Backward Sweep Optimal Power Flow}
\acrodef{FBS}[FBS]{Forward Backward Sweep}
\acrodef{IP}[IP]{Interior Point}
\acrodef{LP}[LP]{Linear Programming}
\acrodef{SOCP}[SOCP]{Second Order Cone Programming}
\acrodef{SDP}[SDP]{Semi Definite Programming}
\acrodef{etaload}[$\eta_{\mathrm{load}}$] {charge efficiency}
\acrodef{etagen}[$\eta_{\mathrm{gen}}$] {discharge efficiency}
\acrodef{etabat}[$\eta_{\mathrm{bat}}$] {battery efficiency}
\acrodef{etain}[$\eta_{\mathrm{in}}$] {converter efficiency}
\acrodef{etatot}[$\eta_{\mathrm{tot}}$] {total battery system efficiency}
\acrodef{ploss}[$P^{\mathrm{loss}}_\mathrm{bat}$]{battery loss power}
\acrodef{pmax}[$P_{\mathrm{max}}$]{max battery power}
\acrodef{R}[$R$] {internal resistance}
\acrodef{cw}[$c_{\mathrm{w}}$] {charge well factor}
\acrodef{SOC}[SOC]{state of charge}
\acrodef{cr}[$c_{\mathrm{r}}$] {recovery factor}
\acrodef{Q}[$Q_{\mathrm{bat}}$] {charge capacity}
\acrodef{C}[$C_{\mathrm{bat}}$] {energy capacity}
\acrodef{Vocavg}[$\bar{V}_{\mathrm{oc}}$] {average open circuit potential}
\acrodef{Vocsoc}[$V_{\mathrm{oc}}$] {average open circuit potential}
\acrodef{Is}[$I_{\mathrm{s}}$]{side current}
\acrodef{vt}[$V_{\mathrm{t}}$]{terminal voltage}
\acrodef{ibat}[$I_{\mathrm{bat}}$]{battery current}
\acrodef{pbat}[$P_{\mathrm{bat}}$]{battery power}
\acrodef{cost}[$c_\mathrm{inv}$]{investment cost}
\acrodef{x}[$x$]{state variable for non-linear and linear battery models. Corresponds to SOC}
\acrodef{x_ex}[$\vec{x}$]{state vector containing $[x_1,x_2]^T$ for the extended non-linear and linear models.}
\acrodef{a}[$a$]{SOC target}
\acrodef{b}[$b$]{cost parameter SOC}
\acrodef{c}[$c$]{cost parameter $u_{\mathrm{gen}}$}
\acrodef{d}[$d$]{cost parameter $u_{\mathrm{load}}$}
\acrodef{e}[$e$]{cost parameter $u_{\mathrm{load}}^2$}
\acrodef{cnet}[$c_{\mathrm{net}}$]{net tariff}
\acrodef{cinv}[$c_{\mathrm{inv}}$]{investment cost}
\acrodef{ncell}[$n_{\mathrm{cell}}$]{number of DUALFOIL cells in series}
\acrodef{Acell}[$A_{\mathrm{cell}}$]{cell area}
\acrodef{dclink}[$v_{\mathrm{DC}}$]{DC link voltage}
\acrodef{pthreshold}[$u_{\mathrm{gen,net}}^{\mathrm{max}}$]{peak shave threshold}
\acrodef{horizonT}[$H_{\mathrm{t}}$]{time horizon}
\acrodef{updateT}[$R_{\mathrm{t}}$]{receding horizon window}
\acrodef{sampleTime}[$T_{\mathrm{s}}$]{sample rate}
\acrodef{rsa}[$r_{\mathrm{sa}}$]{side reaction rate constant anode}
\acrodef{rsc}[$r_{\mathrm{sc}}$]{side reaction rate constant cathode}
\acrodef{unet}[$u_{\mathrm{gen}}^{\mathrm{net}}$]{}
\acrodef{unetmax}[$u_{\mathrm{gen,max}}^{\mathrm{net}}$]{}
\acrodef{ubatgen}[$u_{\mathrm{gen}}^{\mathrm{bat}}$]{}
\acrodef{ubatload}[$u_{\mathrm{load}}^{\mathrm{bat}}$]{}
\acrodef{uloadG2}[$P_{\mathrm{load}}^{\mathrm{G2}}$]{G2 standard industrial load profile}
\acrodef{Qs}[$Q_{\mathrm{s}}$]{lost charge}

\acrodef{ID}[ID]{Identification}
\acrodef{NLS}[NLS]{Nonlinear Least Squares}
\acrodef{LS}[LS]{Least Squares}
\acrodef{MPC}[MPC]{Model Predictive Control}
\acrodef{QP}[QP]{Quadratic Programming}
\acrodef{MSE}[MSE]{Mean Squared Error}
\acrodef{RMSE}[RMSE]{Root Mean Squared Error}
\acrodef{NRMSE}[NRMSE]{Normalized Root Mean Squared Error}
\acrodef{PWA}[PWA]{Piece-Wise Affine}
\acrodef{DOD}[DOD]{Depth Of Discharge}
\acrodef{PFC}[PFC]{Primary Frequency Control}
\acrodef{LFC}[LFC]{Load Frequency Control}
\acrodef{MIQP}[MIQP]{Mixed Integer Quadratic Programming}
\acrodef{MINLP}[MINLP]{Mixed Integer Non Linear Programming}
\acrodef{FC}[FC]{Frequency Control}
\acrodef{SOS}[SOS]{Special Ordered Set}
\acrodef{PV}[PV]{photovaltaic}
\acrodef{ARX}[ARX]{AutoRegressive with eXogenous input}
\acrodef{DSO}[DSO]{Distribution System Operator}
\acrodef{PWA}[PWA]{piecewise-affine}
\acrodef{MILP}[MILP]{Mixed Integer Linear Programming}
\acrodef{SOS}[SOS]{Special Ordered Set}
\acrodef{RT}[RT]{real-time}
\acrodef{SoE}[SoE]{State of Energy}
\acrodef{RHC}[RHC]{Receding Horizon Control}
\acrodef{RPC}[RPC]{Reactive Power Control}
\acrodef{APC}[APC]{Active Power Curtailment}
\acrodef{EoL}[EoL]{End of Life}
\acrodef{NPV}[NPV]{Net Present Value}
\acrodef{IRR}[IRR]{Internal Rate of Return}
\acrodef{ROI}[ROI]{Return on Investment}
\acrodef{RES}[RES]{Renewable Energy Sources}
\acrodef{DG}[DG]{Distributed Generation}
\acrodef{DBS}[DBS]{Distributed Battery Storage}
\acrodef{SoC}[SoC]{State of Charge}

\maketitle

\begin{abstract}
In this paper we present a novel methodology for leveraging Receding Horizon Control (RHC), also known as Model Predictive Control (MPC) strategies for distributed battery storage in a planning problem using a Benders decomposition technique. Longer prediction horizons lead to better storage placement strategies but also higher computational complexity that can quickly become computationally prohibitive. The MPC strategy proposed here in conjunction with a Benders decomposition technique effectively reduces the computational complexity to a manageable level. We use the CIGRE low voltage (LV) benchmark grid as a case study for solving an optimal placement and sizing problem for different control strategies with different MPC prediction horizons. The objective of the MPC strategy is to maximize the photovoltaic (PV) utilization and minimize battery degradation in a local residential area, while satisfying all grid constraints. For this case study we show that the economic value of battery storage is higher when using MPC based storage control strategies than when using heuristic storage control strategies, because MPC strategies explicitly exploit the value of forecast information. The economic merit of this approach can be further increased by explicitly incorporating a battery degradation model in the MPC strategy.    
\end{abstract}


\section*{Nomenclature}
\addcontentsline{toc}{section}{Nomenclature}
\begin{IEEEdescription}[\IEEEsetlabelwidth{$\vec{A}_\mathrm{deg}^u ,\vec{A}_\mathrm{deg}^D$} \parsep=2pt]
	
		\item[$\alpha$]   proxy subproblem costs
		\item[$\alpha_\mathrm{down}$]   lower bound of proxy subproblem costs
		\item[$\vec{\lambda}^{[j]}$]  subproblem dual vector 
		\item[$\vec{\lambda}_\mathrm{s}$]  weighted dual vector 
		\item[$\eta_{\mathrm{dis},i},\eta_{\mathrm{ch},i}$] battery discharging and charging efficiency  
		%
		\item[$\vec{a}_1,\vec{a}_2,\vec{a}_3$]  degradation plane parameter vectors  
		\item[$\vec{A}^{[j]}$] partitioned subproblem inequality matrix 
		\item[\smash{\begin{IEEEeqnarraybox*}[][t]{l}
				\vec{A}_\mathrm{deg}^u, \vec{A}_\mathrm{deg}^D,\\
				\hphantom{\vec{A}_\mathrm{deg}^u,{}} \vec{A}_\mathrm{deg}^z
			\end{IEEEeqnarraybox*}}] matrices to include battery degradation for multiple battery systems and time steps
		\item[$\vec{A}_\mathrm{cost}^{x},\vec{A}_\mathrm{cost}^{y}$] generator cost matrices for single shot problem
		\item[$\tilde{\vec{A}}_\mathrm{cost}^{x},\tilde{\vec{A}}_\mathrm{cost}^{y}$] generator cost matrices for the multiperiod OPF problem 
		\item[$\vec{A}_\mathrm{g}^\mathrm{eq},\vec{A}_\mathrm{g}^\mathrm{in}$]  grid matrices for single shot problem
		\item[$\tilde{\vec{A}}_\mathrm{g}^\mathrm{eq},\tilde{\vec{A}}_\mathrm{g}^\mathrm{in}$] grid matrices for the multiperiod OPF problem
		\item[$\vec{A}_q$]   matrix to describe polygonal P,Q regions 
		\item[$\vec{A}_\mathrm{s}$]  intertemporal storage coupling matrix
		%
		\item[$\vec{b}^\mathrm{c}_i$] offset vector for generator cost segments 
		\item[$\vec{b}^{[j]}$] partitioned subproblem right hand vector 
		\item[$\vec{b}_\mathrm{deg}$]  column vector to include battery degradation 
		\item[$\vec{b}_\mathrm{cost}$]  cost offset vector for the single shot problem 
		\item[$\tilde{\vec{b}}_\mathrm{cost}$]  cost offset vector for the multiperiod problem
		\item[$\vec{b}_\mathrm{g}^\mathrm{eq},\vec{b}_\mathrm{g}^\mathrm{in}$]  grid related column vectors for the single shot problem
		\item[$\tilde{\vec{b}}_\mathrm{g}^\mathrm{eq},\tilde{\vec{b}}_\mathrm{g}^\mathrm{in}$]  grid related column vectors for the multi-period problem 
		\item[$\vec{b}$] loss plane offset vector
		\item[$\vec{b}_\mathrm{s}$] storage coupling right hand vector 
		\item[$\vec{B}$] battery system control input matrix
		\item[$\vec{B}_\mathrm{r}$]  branch flow matrix
		\item[$\vec{B}_\mathrm{v}$] linearized active and reactive power to voltage matrix 
	    \item[$\vec{B}_q$] matrix to describe polygonal P,Q regions	
		%
		\item[$c$] update cycle 
		\item[$c_i^\mathrm{net}$] net power costs and feed-in tariff in \euro/MWh 
		\item[$c_1^\mathrm{pv}$]  PV generator costs in \euro/MWh
		\item[$c_1^\mathrm{s}$]   battery generator costs in \euro/MWh
		\item[$\vec{c}_i$] gradient vector for generator cost segments 
		\item[$\vec{c}_\mathrm{s}$] equivalent battery cost vector in \euro/kWh
		\item[$\vec{c}_\mathrm{d}$]  battery cost vector in \euro/kWh 
		\item[$\vec{C}_\mathrm{g}$]  controllable generator to bus mapping matrix 
		\item[$\vec{C}_\mathrm{pv}$]  PV generator to decisioin variable matrix 
		\item[$\vec{C}_\mathrm{s}$]  battery to decision variable mapping matrix 
		%
		\item[$\vec{d}_k \in \vec{D}$] decision vector for battery degradation
		%
		\item[$E_\mathrm{ld}$]  yearly energy consumption
		\item[$E_\mathrm{net}^{\mathrm{im}}$] yearly imported energy from the feeder
		\item[$\vec{e}$]  state of energy vector 
		\item[$\vec{e}(0)$]  initial state of energy vector	
		\item[$\vec{E}$]  state of energy evolution vector 
		%
		
		%
		
		%
		\item[$H$] control horizon 
		%
		\item[$\vec{i}_\mathrm{b}$]  branch current vector in p.u. 
		\item[$\vec{i}_\mathrm{b}^{\mathrm{max}}$] max branch current vector in p.u. 
		\item[$\vec{i}^0,\vec{i}^1$] supporting current vectors for piecewise linear loss approximation 
		%
		\item[$j$] subproblem index
		\item[$J_\mathrm{sub}$]  sum of subproblem objective values
		\item[$J_\mathrm{sub}^{\mathrm{wS}}$]  yearly revenue with battery storage 
		\item[$J_\mathrm{sub}^{\mathrm{w/oS}}$]  yearly revenue without battery storage 
		\item[$J^{[j]}$] subproblem objective value 
		%
		\item[$k$] time step
		%
		\item[$l$] Benders stage
		\item[$\vec{L}_0,\vec{L}_1$]  supporting plane matrices for piecewise linear loss approximation
		%
		\item[$m$] battery lifetime in years
		\item[$\vec{M}_\mathrm{f}$]  bus-injection to branch-current matrix
		\item[$\vec{M}$]  reduced bus-injection to branch-current matrix 
		%
		\item[$n$] number of subproblems
		\item[$n_\mathrm{b}$] number of buses
		\item[$n_\mathrm{c}$] number of linear constraints
		\item[$n_\mathrm{d}$] number of decision variables
		\item[$n_\mathrm{g}$] number of controllable generators
		\item[$n_\mathrm{l}$] number of branches
		\item[$n_\mathrm{pv}$] number of PV units
		\item[$n_\mathrm{p}$] number of planes
		\item[$n_s$] number of battery systems
		\item[$N$] investment horizon
		
		%
		
		%
		\item[$p_\mathrm{bat}$] active battery stack power
		\item[$\vec{p}$]  nodal active bus power vector 
		\item[$\vec{p}_\mathrm{d},\vec{q}_\mathrm{d}$]  nodal active and reactive power load vectors
		\item[$\vec{p}_\mathrm{gen},\vec{q}_\mathrm{gen}$] active and reactive generator power vectors 
		\item[\smash{\begin{IEEEeqnarraybox*}[][t]{l}\vec{p}_\mathrm{gen}^{\mathrm{pv}},\vec{q}_\mathrm{gen}^{\mathrm{pv}},\\ 
				\hphantom{\vec{p}_\mathrm{gen}^{\mathrm{pv}},{}}\hat{\vec{p}}_\mathrm{gen}^{\mathrm{pv}}\end{IEEEeqnarraybox*}}]  active and reactive PV generator power and prediction vectors 
		\item[$\vec{p}_\mathrm{gen}^{\mathrm{s,dis}}$]  active discharging battery grid power vector 
		\item[$\vec{p}_\mathrm{gen}^{\mathrm{s,ch}} $]  active charging battery grid power vector 
		\item[$\vec{p}_\mathrm{l}^{\mathrm{p}},\vec{p}_\mathrm{l}^{\mathrm{q}}$] decision vectors of real network losses 
		\item[$\vec{p}_\mathrm{d},\vec{q}_\mathrm{d},\hat{\vec{p}}_\mathrm{d}$]  active and reactive power load measurement and prediction vectors 
		\item[$\vec{p}_\mathrm{min}, \vec{p}_\mathrm{max}$] min and max active generator power vectors 
		%
		\item[$\vec{q}$]  nodal reactive bus power vector 
		%
		\item[$\vec{R}_\mathrm{d}$] diagonal branch resistance matrix in p.u. 
		%
		\item[$\vec{s}_\mathrm{max}$] max apparent generator power vector 
		\item[$\vec{S}_x,\vec{S}_u$]  matrices to describe the storage evolution
		%
		\item[$T$] sample time
		%
		%
		\item[$\vec{v}$]  nodal line to neutral RMS voltage vector
		\item[$\vec{v}_\mathrm{min}, \vec{v}_\mathrm{max}$]  min and max nodal RMS voltage vectors 
		\item[$\vec{v}_\mathrm{s}$]  slack bus voltage vector 
		\item[$\vec{V}_\mathrm{df}$] inverse diagonal voltage matrix 
		%
		%
		\item[$\vec{x}_k \in \vec{X}$]   decision vector for grid variables 
		\item[$\vec{X}_\mathrm{d}$]  diagonal branch reactance matrix in p.u.
		%
		\item[$\vec{y}_k \in \vec{Y}$] generator cost decision vector 
		%
		\item[$\vec{z}$]  decision vector for battery capacity variables 
		\item[$\vec{z}_\mathrm{max}$]  upper bound for battery capacity variables	
		\item[$Z_\mathrm{down}$]  lower bound for total profit
		\item[$Z_\mathrm{up}$] upper bound for total profit
\end{IEEEdescription}

\newpage

\section{Introduction}
\label{sec:intro}
\subsection{Motivation}
\IEEEPARstart{E}{nergy} storage technologies can have a key role for decarbonizing the power sector \cite{DeSisternes2016}. In particular, \ac{DBS} in \ac{LV} grids is considered to be a promising technology for balancing short-term fluctations and for alleviating grid congestion caused by a high share of distributed \ac{PV} units \cite{IPCC,Divya2009}. Since a large number of PV units is installed in \ac{LV} grids and since \ac{LV} grid capacity is typically limited, it can be expected that \acp{DSO} will have to increasingly curtail PV output to mitigate grid congestion. This is already the case in Germany, where more than half of the installed PV power, i.e. 22~GW of 39~GW (2015), is installed in \ac{LV} grids \cite[p.~6]{bundesnetzagentur} and where PV curtailment has risen more than tenfold in recent years (2012--2015) \cite[p.~12]{bundesnetzagentur}. With more and more roof-top \ac{PV} units being currently installed in the~US, this issue is likely to attract more attention there as well. In this context, \ac{DBS} configurations can help \acp{DSO} to reduce PV curtailment in \ac{LV} grids. But \ac{DBS} installations in combination with \ac{PV} are also able to increase the self-consumption of PV, thereby lowering electricity costs for end-consumers. This is particularly attractive in net-metering tariff schemes. The authors in \cite{Wogrin2015} report that the optimal size and location of storage depend highly on congested lines in transmission networks and hence on the grid topology. This also holds for \ac{DBS} applications in \ac{LV} grids, where an optimal \ac{DBS} scheme can give support for specific lines that would be overloaded under high \ac{PV} penetration levels.

Furthermore, previous research has drawn attention to \ac{MPC} strategies that allow for an optimal predictive dispatch of energy storage in \ac{LV} grids, for instance \cite{Xu2013,Almassalkhi2015}, in combination with other generation sources. Such strategies solve a multi-period \ac{OPF} problem to optimally schedule generator setpoints over a receding control horizon taking storage dynamics and grid constraints into account. Please note that the application of MPC to power system (dispatch) problems is a long-standing concept, see for example~\cite{Camponogara2002,Venkat2008}. Varying the prediction horizon of the predictive dispatch optimization has a great influence on how much energy can be shifted within this time window. This implies that the optimal energy storage size heavily depends on the employed operational policy (for an illustrative example cf.~\cite[p.~203--205]{Ulbig_Diss2014}). Hence, there is a clear need to combine grid planning considerations with grid operational aspects. 

The main objective of this paper is to develop a planning strategy that leverages \ac{MPC} control strategies acting on different control horizons to find the optimal location and size of \acs{DBS} in \ac{LV} grids.
  
\subsection{Related Work}
Recent studies \cite{Denholm2013,DeSisternes2016,Harsha2011,Harsha2014} evaluate the economic value of energy storage without considering the grid topology. In contrast, the papers \cite{Pandzic2014,Thrampoulidis2015,Wogrin2015,Atwa2010,Castillo2013,Bose2012} incorporate a grid model. They solve a multi-period \ac{OPF} problem over a fixed finite control horizon e.g. 24~hours to obtain the optimal location and size of the storage devices and do not consider the influence of seasonality. By solving a finite horizon problem, the optimal siting results would be different for different seasons, i.e. in summer due to high PV irradiation storage capacities would be bigger than in winter. According to \cite{Pandzic2014} enlarging the horizon makes the siting problem computationally intractable. For this reason, \cite{Pandzic2014} proposes a heuristic to find the best storage locations and capacities by comparing a sequence of multi-period problems resulting in a near-optimal allocation of \ac{DBS} or \cite{Atwa2010} solves a finite problem over four representative days from each season following that the variability within each season cannot be captured. Moreover, considering long horizons does not reflect the operational strategy, since in reality predictive dispatch methods can only act on shorter horizons e.g. in the presence of predictive \ac{RHC} strategies. This is due to the fact that accurate weather and load predictions are only available for limited periods in advance. 

The methods from \cite{Harsha2011,Harsha2014,Pandzic2014,Thrampoulidis2015,Wogrin2015} can deal with uncertainty of PV infeed and load consumption by either solving consecutively-cycled multi-period problems over the course of a reference year \cite{Pandzic2014,Thrampoulidis2015}, a stochastic dynamic programming problem \cite{Harsha2011,Harsha2014}, or for a defined set of \ac{PV} and load scenarios \cite{Wogrin2015}.

To further reduce the computational complexity, \cite{Pandzic2014,Thrampoulidis2015,Wogrin2015} include the so-called DC power flow approximation. Unfortunately, a DC power flow approximation is not applicable for \ac{LV} grids, since active power flow is determined by voltage magnitude differences and not by voltage angle differences as it would be the case in transmission grids. The authors in \cite{Castillo2013,Bose2012} incorporate a semi-definite power flow relaxation, which is still hard to solve. The authors of \cite{Dvijotham2014} incorporate the operational strategy inside a sub-optimal greedy heuristic algorithm that determines the optimal storage size and location. Another way to tackle such complex problems is to decompose the problem into subproblems. This is done in \cite{Nasrolahpour2016,Nick2015} by either using Benders decomposition \cite{Nasrolahpour2016} or via an alternate direction method of multipliers \cite{Nick2015}. In addition, none of the papers include a battery degradation objective in the operational strategy to take battery lifetime into account.

\subsection{Contribution} 
The main contribution of this paper is two-fold. First, we develop a Benders decomposition algorithm for \ac{DBS} applications that links the operational domain with the planning domain. Our approach decomposes with respect to time the sizing and placement problem into a tractable master problem and subproblems. This allows us to account for PV and load uncertainty in the same way as proposed in \cite{Pandzic2014} by incorporating different PV and load realizations inside the coupled subproblems that can be considered as different scenarios. Second, to demonstrate the usefulness of our approach, we present a cost analysis that assesses at which storage cost levels and under which operational strategy a \ac{DBS} investment becomes viable. In our previous work \cite{fortenbacherPSCC16} we developed a linearized \ac{OPF} scheme for \ac{LV} grids and incorporated the resulting \ac{LP} problem in a multi-period \ac{OPF} problem to solve an optimal placement and sizing problem for an infinite control horizon. As an extension we incorporate our linearized \ac{OPF} scheme into an \ac{MPC} control strategy that reflects the operational strategy. The objective of the \ac{MPC} strategy is to maximize the \ac{PV} utilization, while taking battery degradation into account and complying with grid constraints in a local residential area. Unlike~\cite{Baringo2012} the subproblems cannot be solved in a parallel fashion, since storage induces an inherent intertemporal coupling between the subproblems. Nevertheless, we show how we can formulate a Benders decomposition algorithm for this problem class, which significantly reduces the computational effort.

The remainder of this paper is organized as follows. Section~\ref{sec:problemDefinition} defines the problem that we aim to solve. Section~\ref{sec:method} reviews the optimal placement and sizing problem and describes the proposed Benders decomposition method. Section~\ref{sec:results} presents the simulation results and an economic assessment. Finally, Section~\ref{sec:conclusion} presents the conclusions.
\section{Problem Definition}
\label{sec:problemDefinition}
\begin{figure}[!t]
	\centering
	\def\svgwidth{\columnwidth}
	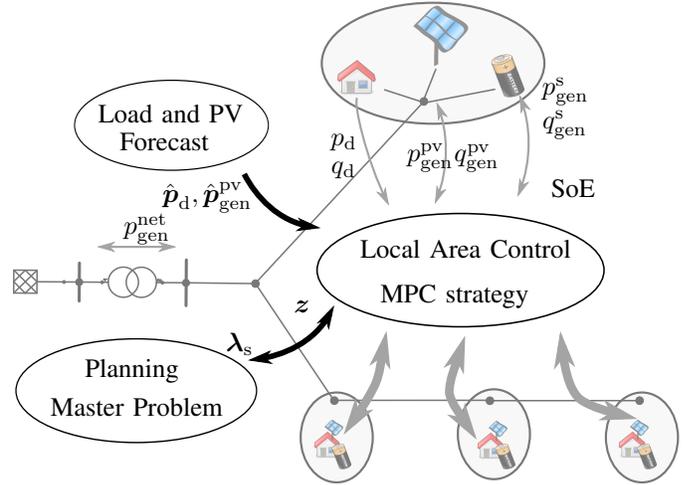
	\caption{Illustration of the local control area. The overall objective of the proposed MPC strategy is to maximize PV utilization and to minimize battery degradation. Dual variables and decision variables are exchanged between planning master problem and MPC storage control strategy.}
	\label{fig:problemDef}
\end{figure}
This section defines the problem that we aim to solve. Figure \ref{fig:problemDef} illustrates the test case environment. We assume that we have a residential local control area with a group of so-called prosumers that have installed \ac{PV} generators and battery systems. The energy capacities $\vec{z}$ are determined by the planning master problem. The area has a centralized local area control entity and a communication infrastructure. The storage control entity is an \ac{MPC} controller that incorporates a multi-period distribution-level \ac{OPF} and acts as a scheduler with control horizons ranging from 24 hours to 1~month. By using an \ac{OPF} method, we enable the optimal utilization of the grid and make use of optimal \ac{APC} and \ac{RPC}. We solve a multi-period problem considering a two-tariff price scenario and a feed-in tariff for the net power~$p_\mathrm{gen}^\mathrm{net}$. In this way, we impose that the scheduler exploits the price differences from the tariff scheme and maximizes the self-consumption of PV in the local area. In addition, we incorporate a battery degradation objective to assess when revenue benefits outweigh degradation cost. The \ac{MPC} controller gets perfect load $\hat{\vec{p}}_\mathrm{d}$ and solar $\hat{\vec{p}}_\mathrm{gen}^\mathrm{pv}$ forecast time-series and schedules the control inputs of the real and reactive powers for the batteries $p_\mathrm{gen}^\mathrm{s},q_\mathrm{gen}^\mathrm{s}$ and the \ac{PV} generators $p_\mathrm{gen}^\mathrm{pv},q_\mathrm{gen}^\mathrm{pv}$. It has also the knowledge of the \ac{SoE} of the batteries acting as a feedback signal to run a new optimization cycle. To account for energy, the \ac{SoE} definition is needed. It differs from the \ac{SoC} due to the nonlinear relationship between open-circuit potential and charge. The dual variables $\vec{\lambda}_\mathrm{s}$ need to be exchanged with the planning problem to initiate a new iteration of the Benders decomposition.

\section{Optimal Placement and Sizing Problem}
\label{sec:method}

\subsection{Linearized \ac{OPF} Problem}
In our previous work \cite{fortenbacherPSCC16} we developed a linearized \ac{OPF} method for radial \ac{LV} networks. Here, we extend our method to incorporate any convex generator cost function and summarize the linear approximations for voltage, branch flow, and network power losses from \cite{fortenbacherPSCC16}.

\subsubsection{Voltage Approximation} Under the assumption of a high $R/X$ ratio, the absolute voltage magnitude drops along $n_\mathrm{l}$ lines in a radial balanced grid are linearly approximated by  
\begin{equation}
\vec{v}-\vec{v}_\mathrm{s} \approx \underbrace{\left[ \vec{M}^T\vec{R}_\mathrm{d}\vec{M}_\mathrm{f}{\vec{V}}_\mathrm{df} \quad \vec{M}^T\vec{X}_\mathrm{d}\vec{M}_\mathrm{f}{\vec{V}}_\mathrm{df} \right]}_{\vec{B}_\mathrm{v}}\left[\begin{array}{c} \vec{p}  \\
\vec{q}  \end{array} \right] , \label{eq:vapprox}
\end{equation}
\noindent where $\vec{v} \in \mathbb{R}^{n_\mathrm{l}\times 1}$ is the nodal absolute voltage magnitude vector in per unit and $\vec{v}_\mathrm{s} \in \mathbb{R}^{n_\mathrm{l}\times 1}$ is the absolute voltage magnitude vector for the slack bus in per unit. The vectors $\vec{p},\vec{q} \in \mathbb{R}^{n_\mathrm{b} \times 1} $ are the nodal per unit injections for active and reactive power for $n_\mathrm{b}$ buses. The matrix $\vec{M}_\mathrm{f} \in \mathbb{R}^{n_\mathrm{l} \times n_\mathrm{b}}$ maps the bus-injection currents to branch-currents and is also called the bus-injection to branch-current (BIBC) matrix. Here, we also define a reduced version of $\vec{M}_\mathrm{f}$ indicated with $\vec{M} \in \mathbb{R}^{n_\mathrm{l} \times n_\mathrm{b}-1}$, in which the row of the involved slack bus is deleted. $\vec{R}_\mathrm{d}^{n_\mathrm{l} \times n_\mathrm{l}} = \mathrm{diag}\{r_{\mathrm{d}1},...,r_{\mathrm{d}n_\mathrm{l}} \}$ is the branch resistance matrix in per unit and $\vec{X}_\mathrm{d}^{n_\mathrm{l} \times n_\mathrm{l}} = \mathrm{diag}\{x_{\mathrm{d}1},...,x_{\mathrm{d}n_\mathrm{l}} \}$ is the reactance matrix in per unit. The matrix ${\vec{V}}_\mathrm{df} \in \mathbb{R}^{n_\mathrm{b} \times n_\mathrm{b}}$ includes the inverse nodal complex voltages of the grid and is defined as  
\begin{equation}
{\vec{V}}_\mathrm{df} = |\mathrm{diag}\{1/\underline{v}_1,\ldots,1/\underline{v}_{n_\mathrm{b}}\}^*| \quad.
\end{equation}
\subsubsection{Branch Flow Approximation} If we assume that the reactive power injections are much smaller than the active power injections, which holds for normal grid operation in \ac{LV} grids, we can neglect the contribution on the reactive power by approximating 
\begin{equation}
\vec{i}_\mathrm{b} \approx \underbrace{\vec{M}_\mathrm{f}{\vec{V}}_\mathrm{df}}_{\vec{B}_\mathrm{r}}\vec{p}  \quad, \label{eq:branch}
\end{equation}
\noindent where $\vec{i}_\mathrm{b} \in \mathbb{R}^{n_\mathrm{l} \times 1}$ is the current branch flow magnitude vector in per unit. 

\subsubsection{Loss Approximation}
We approximate active network power losses by
\begin{align}
\vec{p}_\mathrm{l}^\mathrm{p} & \approx \max\left\{\vec{L}_0\vec{p},-\vec{L}_0\vec{p},\vec{L}_1\vec{p} +\vec{b},-\vec{L}_1\vec{p} +\vec{b} \right\}\quad, \label{eq:pwalosses} \\
\vec{p}_\mathrm{l}^\mathrm{q} & \approx \max\left\{\vec{L}_0\vec{q},-\vec{L}_0\vec{q},\vec{L}_1\vec{q} +\vec{b},-\vec{L}_1\vec{q} +\vec{b} \right\} \label{eq:epiq} \quad.
\end{align}
\noindent where 
\begin{eqnarray}
\vec{L}_0 &= &\mathrm{diag}\{i^0_{1},\cdots,i^0_{n_\mathrm{l}} \} \vec{R}_\mathrm{d}\vec{M}_\mathrm{f}{\vec{V}}_\mathrm{df} \quad, \label{eq:plane1} \\
\vec{L}_1 & =&\mathrm{diag}\{i^0_{1}+i^1_{1},\cdots,i^0_{n_\mathrm{l}}+i^1_{n_\mathrm{l}}\}\vec{R}_\mathrm{d}\vec{M}_\mathrm{f}{\vec{V}}_\mathrm{df} \quad,\\
\vec{b} &= &-[r_1 i^0_{1}i^1_{1},\cdots, r_{n_\mathrm{l}} i^0_{n_\mathrm{l}}i^1_{n_\mathrm{l}}]^T \label{eq:plane3} \quad. 
\end{eqnarray}
The power line loss vectors $\vec{p}_\mathrm{l}^\mathrm{p} \in \mathbb{R}^{n_\mathrm{l} \times 1}$ and $\vec{p}_\mathrm{l}^\mathrm{q} \in \mathbb{R}^{n_\mathrm{l} \times 1}$ are the real network power losses resulting from active and reactive power injections. Equations \eqref{eq:plane1}-\eqref{eq:plane3} define hyperplanes with the supporting currents $\vec{i}_0,\vec{i}_1$ for the power losses that are inner approximations of the quadratic power loss functions.

\subsubsection{\ac{FBS-OPF} Formulation}
With the presented approximations we can state a linearized \ac{OPF} problem. Since the approximations are in the \ac{FBS} load flow framework \cite{Teng}, we call our linearized method \ac{FBS-OPF}. We first define the optimization vector $\vec{x} = \left[\vec{p}_\mathrm{l}^\mathrm{p},\vec{p}_\mathrm{l}^\mathrm{q},\vec{p}_\mathrm{gen},\vec{q}_\mathrm{gen},\vec{v} \right]^T$ reflecting all grid related variables. The helper decision vector $\vec{y} \in \mathbb{R}^{n_\mathrm{g}\times 1} $ specifies the generation costs for $n_\mathrm{g}$ generators. The objective is to find the optimal generator setpoints $\vec{p}_\mathrm{gen},\vec{q}_\mathrm{gen}$ that minimize the generation costs, while satisfying all grid constraints. The active and reactive generator bus injections $\vec{p}_\mathrm{gen} \in \mathbb{R}^{n_\mathrm{g} \times 1}$ and $\vec{q}_\mathrm{gen} \in \mathbb{R}^{n_\mathrm{g}\times 1 }$  are mapped to the buses with the matrix $\vec{C}_\mathrm{g} \in \mathbb{R}^{n_\mathrm{b} \times n_\mathrm{g}}$. 

\begin{figure}[!t]
	\centering
		\def\svgwidth{\columnwidth}
	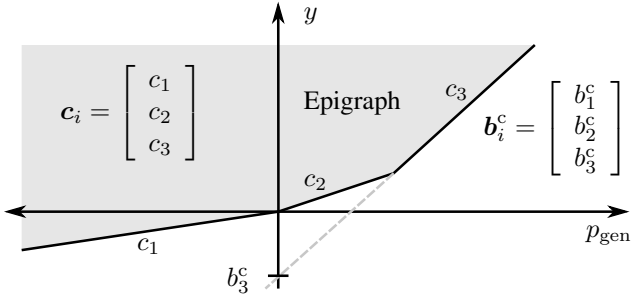
	\caption{Piecewise-affine (PWA) cost function representation showing illustratively three line segments to reflect any convex generator cost curves.}
	\label{fig:pwafunction}
\end{figure}

\begin{figure}[!t]
	\centering
	\def\svgwidth{\columnwidth}
	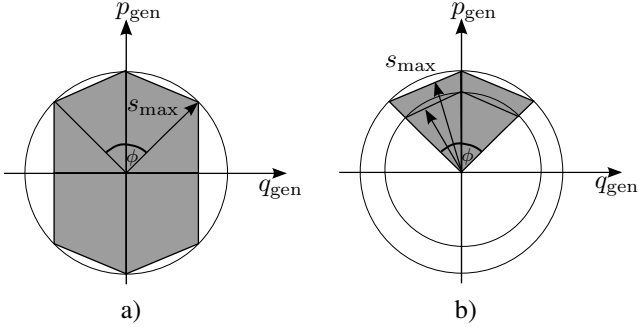
	\caption{Approximated reactive power capability areas a) circular-bounded b) $\cos\phi$-bounded. The polygonal convex regions can be described with the constraints (\ref{eq:FBOPF}h-l). \cite{fortenbacherPowerTech2015}}
	\label{fig:reactive}
\end{figure}

The extended linearized optimization problem is then:
\begin{equation}
\begin{array}{lllll}
\multicolumn{5}{l}{J^*=\displaystyle\min_{\vec{x},\vec{y}} \vec{1}^T \vec{y} } \\
& \text{s.t.} \\
&\text{(a)}&   \left[\begin{array}{@{\hspace{2pt}}c@{\hspace{2pt}}c@{\hspace{2pt}}c@{\hspace{2pt}}} \vec{c}_1 & & \\ & \ddots & \\ & & \vec{c}_{n_{\mathrm{g}}}\end{array} \right] \vec{p}_\mathrm{gen} - \left[\begin{array}{@{\hspace{2pt}}c@{\hspace{2pt}}c@{\hspace{2pt}}c@{\hspace{2pt}}} \vec{1} & & \\ & \ddots & \\ & & \vec{1}\end{array} \right] \vec{y} \leq - \left[\begin{array}{c} \vec{b}^\mathrm{c}_1 \\ \vdots \\ \vec{b}^\mathrm{c}_{n_\mathrm{g}}  \end{array}\right] \\   
& \text{(b)} & \multicolumn{3}{l}{\vec{1}^T\vec{C}_\mathrm{g}\vec{p}_\mathrm{gen} - \vec{1}^T\vec{p}_\mathrm{l}^\mathrm{p}-\vec{1}^T\vec{p}_\mathrm{l}^\mathrm{q}  = \vec{1}^T\vec{p}_\mathrm{d}  }\\
& \text{(c)} &  \multicolumn{3}{l}{\vec{B}_\mathrm{v} \left[\begin{array}{l} \vec{C}_\mathrm{g}\vec{p}_\mathrm{gen} \\ \vec{C}_\mathrm{g}\vec{q}_\mathrm{gen}  \end{array} \right] -\vec{v}  = \vec{B}_\mathrm{v}\left[\begin{array}{l} \vec{p}_\mathrm{d} \\ \vec{q}_\mathrm{d}  \end{array} \right] - \vec{v}_\mathrm{s} } \\
& \text{(d)} & \multicolumn{3}{l}{\vec{p}_\mathrm{l}^\mathrm{p} - \vec{L}_0\vec{C}_\mathrm{g}\vec{p}_\mathrm{gen} \geq -\vec{L}_0\vec{p}_\mathrm{d} } \\
& \text{(e)} & \multicolumn{3}{l}{\vec{p}_\mathrm{l}^\mathrm{p} + \vec{L}_0\vec{C}_\mathrm{g}\vec{p}_\mathrm{gen} \geq \vec{L}_0\vec{p}_\mathrm{d} } \\
& \text{(f)} & \multicolumn{3}{l}{\vec{p}_\mathrm{l}^\mathrm{p} - \vec{L}_1\vec{C}_\mathrm{g}\vec{p}_\mathrm{gen} \geq -\vec{L}_1\vec{p}_\mathrm{d} + \vec{b} } \\
& \text{(g)} & \multicolumn{3}{l}{\vec{p}_\mathrm{l}^\mathrm{p} + \vec{L}_1\vec{C}_\mathrm{g}\vec{p}_\mathrm{gen} \geq +\vec{L}_1\vec{p}_\mathrm{d} + \vec{b} } \\
& \text{(h)} & \multicolumn{3}{l}{\vec{p}_\mathrm{l}^\mathrm{q} - \vec{L}_0\vec{C}_\mathrm{g}\vec{q}_\mathrm{gen} \geq -\vec{L}_0\vec{q}_\mathrm{d} } \\
& \text{(i)} & \multicolumn{3}{l}{\vec{p}_\mathrm{l}^\mathrm{q} + \vec{L}_0\vec{C}_\mathrm{g}\vec{q}_\mathrm{gen} \geq \vec{L}_0\vec{q}_\mathrm{d} } \\
& \text{(j)} & \multicolumn{3}{l}{\vec{p}_\mathrm{l}^\mathrm{q} - \vec{L}_1\vec{C}_\mathrm{g}\vec{q}_\mathrm{gen} \geq -\vec{L}_1\vec{q}_\mathrm{d} + \vec{b} } \\
& \text{(k)} & \multicolumn{3}{l}{\vec{p}_\mathrm{l}^\mathrm{q} + \vec{L}_1\vec{C}_\mathrm{g}\vec{q}_\mathrm{gen} \geq +\vec{L}_1\vec{q}_\mathrm{d} + \vec{b} } \\
& \text{(l)} & \multicolumn{3}{l}{-\vec{i}_\mathrm{b}^\mathrm{max} + \vec{B}_\mathrm{r}\vec{p}_\mathrm{d}\leq \vec{B}_\mathrm{r}\vec{C}_\mathrm{g}\vec{p}_\mathrm{gen} \leq \vec{i}_\mathrm{b}^\mathrm{max} + \vec{B}_\mathrm{r}\vec{p}_\mathrm{d}}\\
& \text{(m)} &\vec{v}_\mathrm{min} \leq \vec{v} \leq \vec{v}_\mathrm{max} \\
& \text{(n)} &\vec{p}_\mathrm{min} \leq \vec{p}_\mathrm{gen} \leq \vec{p}_\mathrm{max} \\
& \text{(o)} & -\vec{s}_{\mathrm{max}} \leq \vec{p}_\mathrm{gen} +  \vec{A}_q\vec{q}_\mathrm{gen} \leq \vec{s}_{\mathrm{max}} \\
& \text{(p)} & -\vec{s}_{\mathrm{max}} \leq \vec{p}_\mathrm{gen} - \vec{A}_q\vec{q}_\mathrm{gen} \leq \vec{s}_{\mathrm{max}} \\
& \text{(q)} & -\vec{B}_q\vec{s}_{\mathrm{max}} \leq \vec{q}_\mathrm{gen} \leq\vec{B}_q\vec{s}_{\mathrm{max}} \quad,
\end{array}
\label{eq:FBOPF}
\end{equation}
\noindent where $\vec{p}_\mathrm{d} \in \mathbb{R}^{n_\mathrm{b} \times 1}$ and $\vec{q}_\mathrm{d} \in \mathbb{R}^{n_\mathrm{b} \times 1}$ are the active and reactive net load. Constraint  (\ref{eq:FBOPF}a) includes the epigraphs of convex \ac{PWA} generator cost functions, i.e. the set of points lying above the specified cost functions. The vectors $\vec{c}_i \in \mathbb{R}^{n_\mathrm{ls}\times 1}$ and $\vec{b}^\mathrm{c}_i \in \mathbb{R}^{n_\mathrm{ls}\times 1}$ assign the gradients and offsets of the \ac{PWA} cost function for each generator. The variable $n_\mathrm{ls}$ specifies the number of the cost function segments. In this way, we can model the generation costs as a set of linear constraints. Figure~\ref{fig:pwafunction} illustrates the \ac{PWA} cost function representation for three line segments and one generator. Constraint (\ref{eq:FBOPF}b) enforces power balance in the grid. The voltage approximation \eqref{eq:vapprox} is included in (\ref{eq:FBOPF}c). The constraints (\ref{eq:FBOPF}d-k) incorporate epigraph formulations of \eqref{eq:pwalosses} and \eqref{eq:epiq} that are piecewise linear inner approximations of the real power losses. Constraint (\ref{eq:FBOPF}l) includes the branch flow limit approximation \eqref{eq:branch}. Constraints (\ref{eq:FBOPF}m,n) specify the lower and upper bounds for the voltage ($\vec{v}_\mathrm{min}, \vec{v}_\mathrm{max}$) and active generator powers  ($\vec{p}_\mathrm{min}, \vec{p}_\mathrm{max}$). Constraints (\ref{eq:FBOPF}o-q) approximate the generators' apparent power limits, where $\vec{s}_\mathrm{max}$ is the generators' maximum apparent power; specifically, we define circular-bounded and  $\cos\phi$-bounded active and reactive power settings by approximating the circular area/segments with convex sets \cite{fortenbacherPowerTech2015} that describe the polygons depicted in Fig~\ref{fig:reactive}.

For the sake of convenience, problem \eqref{eq:FBOPF} can be written in a more compact form:
\begin{equation}
\begin{array}{lll}
J^{*} = & \multicolumn{2}{l} {\displaystyle\min_{\vec{x,y}} \ \vec{1}^T \vec{y}} \\
& \text{s.t.} \\
&\text{(a)}&  \vec{A}_\mathrm{cost}^{x} \vec{x} - \vec{A}_\mathrm{cost}^{y} \vec{y} \leq -\vec{b}_{\mathrm{cost}}\\
&\text{(b)}&  \vec{A}_\mathrm{g}^\mathrm{in}\vec{x}    \geq \vec{b}_\mathrm{g}^\mathrm{in} \\
& \text{(c)}& \vec{A}_\mathrm{g}^\mathrm{eq}  \vec{x}    = \vec{b}_\mathrm{g}^\mathrm{eq} \\
& \text{(d)}& \vec{x}_\mathrm{min} \leq \vec{x} \leq \vec{x}_\mathrm{max} \quad,
\end{array}
\label{eq:singleshot} 
\end{equation}
\noindent where constraint (\ref{eq:singleshot}a) incorporates (\ref{eq:FBOPF}a), (\ref{eq:singleshot}b) incorporates (\ref{eq:FBOPF}d-l,o-q), (\ref{eq:singleshot}c) incorporates (\ref{eq:FBOPF}b,c), and (\ref{eq:singleshot}d) incorporates (\ref{eq:FBOPF}m,n).
\subsection{Multi-period Problem}
Since energy storage introduces an intertemporal coupling into our dispatch problems, the placement and sizing problem has to be formulated as a multi-period problem over a given investment horizon $N$. By incorporating the single shot solutions of \eqref{eq:singleshot}, the purpose is to find the optimal placement and sizes of storage devices, while considering a certain operational strategy. We introduce a further optimization vector $\vec{z}\in \mathbb{R}^{n_\mathrm{s}\times 1 } $ that specifies the energy capacities of $n_\mathrm{s}$ batteries. Here, we require that battery capacities are continuous variables for complexity reasons. This is based on the assumption that batteries are scalable devices in size. We extend $\vec{X} = \left[\vec{x}_0,\ldots,\vec{x}_{N-1}\right]^T$ and  $\vec{Y} = \left[\vec{y}_0,\ldots,\vec{y}_{N-1}\right]^T$ to account for multiple steps. The sizing and placement problem can be written as follows:
\begin{equation}
\begin{array}{lll}
J^{*} = & \multicolumn{2}{l} {\displaystyle\min_{\vec{X},\vec{Y},\vec{z},\vec{D}} \ \underbrace{T \left(\sum\limits_{k=0}^{N-1} \ \vec{1}^T \vec{y}(k) + \vec{c}_\mathrm{d}^T \vec{d}(k) \right)}_{\text{costs}} + \ \underbrace{\vec{c}_\mathrm{s}^T\vec{z}}_{\substack{\text{storage} \\ {\text{investment}}}} } \\
& \text{s.t.} \\
&\text{(a)}&  \tilde{\vec{A}}_\mathrm{cost}^{x} \vec{X} - \tilde{\vec{A}}_\mathrm{cost}^{y} \vec{Y} \leq -\tilde{\vec{b}}_{\mathrm{cost}}\\
&\text{(b)}&  \tilde{\vec{A}}_\mathrm{g}^\mathrm{in}\vec{X}    \geq \tilde{\vec{b}}_\mathrm{g}^\mathrm{in} \\
& \text{(c)}& \tilde{\vec{A}}_\mathrm{g}^\mathrm{eq}  \vec{X}    = \tilde{\vec{b}}_\mathrm{g}^\mathrm{eq} \\
& \text{(d)}& \vec{A}_\mathrm{s}  \left[ \begin{array}{c}  \vec{X} \\ \vec{z} \end{array} \right]   \leq \vec{b}_\mathrm{s} \vspace{0.1cm}\\
& \text{(e)}& [\vec{A}_\mathrm{deg}^u \ \vec{A}_\mathrm{deg}^z \ \vec{A}_\mathrm{deg}^D] \left[ \begin{array}{c} \vec{X} \\ \vec{z} \\ \vec{D} \end{array}\right]   \leq \vec{b}_\mathrm{deg} \\
& \text{(f)}& \vec{X}_\mathrm{min} \leq \vec{X} \leq \vec{X}_\mathrm{max} \quad,
\end{array}
\label{eq:placesize} 
\end{equation}
\noindent where $\vec{c}_\mathrm{d}$ specifies the total battery cost, $\vec{c}_\mathrm{s}$ is the equivalent battery cost for the given investment horizon $N$ related to the battery lifetime, and $T$ is the sample interval. In contrast to our previous work, we also include battery degradation with $\vec{D} = [\vec{d}(0),\hdots,\vec{d}(N-1) \in \mathbb{R}^{n_s\times 1}]^T$ representing the capacity loss evolution. The overall objective consists of two parts: (1) storage investment and (2) operational costs. Note that the costs can take negative values that would correspond to revenue. Therefore, problem \eqref{eq:placesize} can also be regarded as a profit maximization problem. The constraints of problem \eqref{eq:placesize} are described in the following subsections.

\subsubsection{Generator Cost Functions (\ref{eq:placesize}a)}
Any convex cost structure of the operational domain can be considered by applying different cost data for the individual time steps:
\begin{align}
\tilde{\vec{A}}_\mathrm{cost}^x = &  \text{blkdiag} \{\vec{A}_{\mathrm{cost},0}^x,\ldots,\vec{A}_{\mathrm{cost},N-1}^x \} \ ,\\
\tilde{\vec{A}}_\mathrm{cost}^y = &  \text{blkdiag} \{\vec{A}_{\mathrm{cost},0}^y,\ldots,\vec{A}_{\mathrm{cost},N-1}^y \} \ ,	\\
\tilde{\vec{b}}_{\mathrm{cost}} = & \left[\vec{b}_{\mathrm{cost},0},\ldots,\vec{b}_{\mathrm{cost},N-1} \right]^T \quad.
\end{align}
This generic representation allows us to model various tariff schemes such as high and low tariff schemes in combination with feed-in tariffs or even energy price profiles.

\subsubsection{Grid constraints (\ref{eq:placesize}b,c)}
To comply with the multi-period problem structure, the following matrices need to be replicated for each time step:
\begin{align}
\tilde{\vec{A}}_\mathrm{g}^\mathrm{in} = &  \text{blkdiag} \{\vec{A}_{\mathrm{g},0}^\mathrm{in},\ldots,\vec{A}_{\mathrm{g},N-1}^\mathrm{in} \} \quad,\\
\tilde{\vec{b}}_\mathrm{g}^\mathrm{in} = & [\vec{b}_{\mathrm{g},0}^\mathrm{in},\ldots,\vec{b}_{\mathrm{g},N-1}^\mathrm{in} ]^T\quad, \\
\tilde{\vec{A}}_\mathrm{g}^\mathrm{eq} = & \text{blkdiag} \{  
\vec{A}_{\mathrm{g},0}^\mathrm{eq},\ldots, \vec{A}_{\mathrm{g},N-1}^\mathrm{eq} \} \quad, \\
\tilde{\vec{b}}_\mathrm{g}^\mathrm{eq} = & [\vec{b}_{\mathrm{g},0}^\mathrm{eq},\ldots, \vec{b}_{\mathrm{g},N-1}^\mathrm{eq}]^T \quad.
\end{align}

\subsubsection{Incorporation of Storage (\ref{eq:placesize}d)}
We can define the \ac{SoE} vector $\vec{e} = [e_1,\hdots,e_{n_\mathrm{s}}]^T$ at time step $k$ by
\begin{equation}
\vec{e}(k+1)  = \vec{I}^{n_\mathrm{s}}\vec{e}(k) + 
\vec{B} \left[\begin{array}{l} \vec{p}_\mathrm{gen}^\mathrm{s,dis}(k) \\ \vec{p}_\mathrm{gen}^\mathrm{s,ch}(k) \end{array}\right]  \quad,									
\end{equation}
 \noindent where $\vec{p}_\mathrm{gen}^\mathrm{s,dis}(k) \geq \vec{0} \in \mathbb{R}^{n_s \times 1}$, $\vec{p}_\mathrm{gen}^\mathrm{s,ch}(k) < \vec{0} \in \mathbb{R}^{n_s \times 1}$ are the total discharging and charging powers of the storage units, and $\vec{I}^{n_s}$ denotes the identity matrix of dimension $n_\mathrm{s}$. The input matrix $\vec{B} \in \mathbb{R}^{n_\mathrm{s} \times 2n_\mathrm{s}}$ is
\begin{equation}
\vec{B} =  T \left[-\mathrm{diag}\{\eta_{\mathrm{dis},1}^{-1},...,\eta_{\mathrm{dis},n_\mathrm{s}}^{-1}\}  \ \mathrm{diag}\{\eta_{\mathrm{ch},1},...,\eta_{\mathrm{ch},n_\mathrm{s}}\}  \right] \quad,
\end{equation}
\noindent where $\eta_{\mathrm{ch},i},\eta_{\mathrm{dis},i}$ are the charging and discharging efficiencies. To incorporate the complete energy level evolution $\vec{E} = [\vec{e}(1),...,\vec{e}(N-1)]^T$, we define
\begin{equation}
\vec{E} =  \underbrace{\left[ \begin{array}{c} \vec{I} \\ \vdots \\ \vec{I}\end{array} \right]}_{\vec{S}_x}\vec{e}(0) + \underbrace{\left[\begin{array}{ccc}  \vec{B}\vec{C}_\mathrm{s}  &  & \vec{0} \\ \vdots &\ddots \\ \vec{B}\vec{C}_\mathrm{s} & \cdots & \vec{B}\vec{C}_\mathrm{s} \end{array}\right]}_{\vec{S}_u}\underbrace{\left[\begin{array}{l} \vec{x}_0 \\ \vdots \\  \vec{x}_{N-1} \end{array} \right]}_{\vec{X}} \quad, \label{eq:evo}
\end{equation}
\noindent where  $\vec{e}(0)$ denotes the initial \ac{SoE} vector. The storage power variables for charging and discharging are mapped with the matrix $\vec{C}_\mathrm{s} \in \mathbb{R}^{2n_\mathrm{s}\times 2n_\mathrm{l}+2n_\mathrm{g}+n_\mathrm{b}}$ to the vector $\vec{x}_k$ by
\begin{equation}
\left[\begin{array}{l} \vec{p}_\mathrm{gen}^\mathrm{s,dis}(k) \\ \vec{p}_\mathrm{gen}^\mathrm{s,ch}(k) \end{array}\right] = \vec{C}_\mathrm{s}\vec{x}_k \quad.
\end{equation}

We can specify the following constraint sets
\begin{align}
\vec{A}_\mathrm{s} = & \left[\begin{array}{rc} \vec{S}_u & [-\vec{1}^{N\times 1}\otimes \vec{I}^{n_\mathrm{s}} ]\\ -\vec{S}_u & \vec{0} \end{array} \right] , \vec{b}_\mathrm{s} = & \left[\begin{array}{r}-\vec{S}_{x} \vec{e}(0) \\ \vec{S}_{x} \vec{e}(0) \end{array} \right] \quad,
\end{align}
to define the minimum and maximum energy \ac{SoE} bounds as a function of the variable storage capacities $\vec{z}$ and $\vec{X}$. The operator $\otimes$ defines the Kronecker product.

\subsubsection{Incorporation of Degradation (\ref{eq:placesize}e)}
\begin{figure}[!t]
	\centering
	\includegraphics[width = \columnwidth]{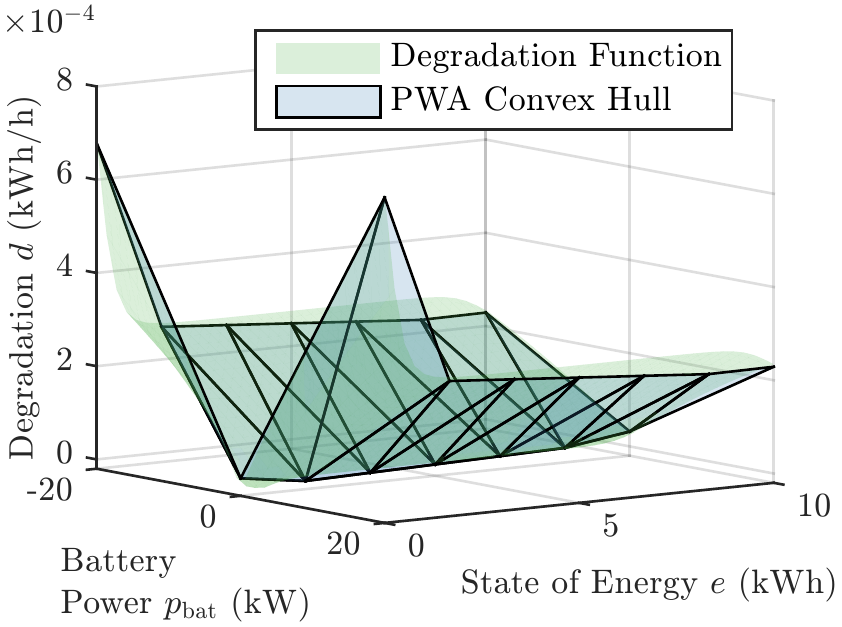}
	\caption{Illustration of a degradation map with an energy capacity of $z=$10kWh showing the incremental capacity loss as a function of the State of Energy (SoE) $e$ and applied battery power $p_\mathrm{bat}$. The red surface is the original degradation function from \cite{Forman2012}. The blue \acf{PWA} map is its convex hull representation \eqref{eq:hull}.}
	\label{fig:degMap}
\end{figure}

Battery degradation reduces the available battery capacity. It has an impact on the overall profitability, since the storage revenue decreases over time due to the capacity loss. There is a complex relationship between battery degradation and operational management, such that the operational strategy has an impact on lifetime and profitability. In \cite{fortenbacherPSCC14}, we presented a method to identify a stationary degradation process on an arbitrary battery usage pattern. The method produces a degradation map, where degradation is a function of the battery power and \ac{SoE}. By using a convex \ac{PWA} representation of the degradation map it is possible to account for degradation in the operational domain with efficient optimization solvers.  As an illustrative example, we show a degradation map for a LiFePO4 system in Fig.~\ref{fig:degMap}. The green surface represents an empirical degradation function from \cite{Forman2012} that we have scaled and transformed to the energy/power domain. The blue \ac{PWA} map denotes its convex hull.

To calculate the total capacity fade per time $d$ e.g. in kWh/h as a function of the \ac{SoE} $e$, battery power $p_\mathrm{bat}$, and energy capacity $z$, the \ac{PWA} map for one battery system has the following structure
\begin{equation}
d =  \max \left(\left[\vec{a}_1 \ \vec{a}_2 \ \vec{a}_3 \right]  \left[\begin{array}[h]{c}
p_{\mathrm{bat}} \\
e \\
z
\end{array}\right]  \right) \quad,
\label{eq:hull}
\end{equation}
\noindent where $\vec{a}_1,\vec{a}_2,\vec{a}_3 \in \mathbb{R}^{n_\mathrm{p}}$ are parameters that span $n_\mathrm{p}$ planes in $\mathbb{R}^3$. 
The incremental capacity loss for one battery system at time step $k$ can be included inside the optimal placement and sizing problem \eqref{eq:placesize} by using the following epigraph formulation:
\begin{equation}
\vec{a}_1 (p_\mathrm{gen}^{s,\mathrm{dis}}(k)+p_\mathrm{gen}^{s,\mathrm{ch}}(k)) + \vec{a}_2 e(k) + \vec{a}_3 z \leq \vec{1} d(k) \quad,
\end{equation}
where $p_\mathrm{bat} = p_\mathrm{gen}^{s,\mathrm{dis}}+p_\mathrm{gen}^{s,\mathrm{ch}}$. In the same straightforward way, we can account for battery degradation for multiple steps and battery systems, the following matrix definitions need to be included into the placement and sizing problem:
\begin{align}
\vec{A}_\mathrm{deg}^u = &[\vec{I}^{Nn_\mathrm{s}} \otimes \vec{a}_1][\vec{I}^{N} \otimes [\vec{I}^{n_\mathrm{s}} \vec{I}^{n_\mathrm{s}}] \vec{C}_\mathrm{s}] + [\vec{I}^{Nn_\mathrm{s}}\otimes \vec{a}_2]\vec{S}_u \ ,\\
\vec{A}_\mathrm{deg}^z = &\vec{I}^{N}\otimes [\vec{I}^{n_\mathrm{s}}\otimes \vec{a}_3] \quad, \\
\vec{A}_\mathrm{deg}^D = &\vec{I}^{N n_\mathrm{s}}\otimes -\vec{1}^{n_\mathrm{p}\times 1} \quad, \\
\vec{b}_\mathrm{deg} = & -[ \vec{I}^{Nn_\mathrm{s}}\otimes \vec{a}_2]\vec{S}_x\vec{e}(0) \quad.
\end{align}

\subsubsection{Incorporation of PV generators (\ref{eq:placesize}f)}
The PV predictions $\hat{\vec{p}}_\mathrm{gen}^{\mathrm{pv}}$ for $n_\mathrm{pv}$ PV generators can be included as time series into constraint (\ref{eq:placesize}f) by applying at each time step $k$
\begin{equation}
\hat{\vec{p}}_\mathrm{gen}^{\mathrm{pv}}(k) = \vec{C}_{\mathrm{pv}}\vec{x}_{\max,k} \quad,
\end{equation}
\noindent where $\vec{C}_\mathrm{pv} \in \mathbb{R}^{n_\mathrm{pv}\times 2n_\mathrm{l}+2n_\mathrm{g}+n_\mathrm{b}}$ maps the PV generators to the optimization vector $\vec{x}_k$.

\subsubsection{Problem Complexity}
The average polynomial running time of solving an \ac{LP} problem with the Simplex method can be approximated according to \cite{schrijver1998theory} by
\begin{equation}
\mathcal{O}\left(n_\mathrm{d}^3n_\mathrm{c}^{1/(n_\mathrm{d}-1)}\right) \quad  ,
\end{equation}
\noindent where $n_\mathrm{d}$ is the size of the decision variables and $n_\mathrm{c}$ denotes the number of constraints. Taking the specific problem structure into account, the complexity bound for solving the \ac{LP} problem~\eqref{eq:placesize} can be calculated as
\begin{equation}
\mathcal{O} \left(n_\mathrm{d}^3\right) , 
\end{equation}
\noindent where the size of $n_\mathrm{d} \approx N(3 n_\mathrm{g} + 2 n_\mathrm{l} + n_\mathrm{b} + n_\mathrm{s})$. Here, it is assumed that for large $N$ the term $n_\mathrm{c}^{1/(n_\mathrm{d}-1)} \rightarrow 1 $. This means that the computation time depends strongly on the investment horizon $N$.

\subsection{Benders Decomposition}
Since the \ac{LP} problem \eqref{eq:placesize} is intractable for infinite control horizons, we try to decompose our placement and sizing problem with respect to time. By exploiting the \ac{LP} property, we can decompose our problem using Benders decomposition. We use the same decomposition procedure and notation according to \cite{Conejo2006}.
\subsubsection{Master Problem}
According to \cite{Conejo2006} problem \eqref{eq:placesize} has a decomposable structure, since $\vec{z}$ acts as the complicating variable. Hence, the problem can be split into a storage planning master problem and sequentially-solvable subproblems reflecting the operational strategy. The master problem is
\begin{equation}
\begin{array}{lll}
J^{*}(J_\mathrm{sub}^{(l)},\vec{z}^{(l)},\vec{\lambda}_\mathrm{s}^{(l)}) = & \multicolumn{2}{l} {\displaystyle\min_{\vec{z},\alpha} \ \vec{c}_\mathrm{s}^T\vec{z} + \alpha } \\
& \text{s.t.} \\
& \text{(a)}&  J_\mathrm{sub}^{(l)} + {\vec{\lambda}_{\mathrm{s}}^{T}}^{(l)}(\vec{z}-\vec{z}^{(l)}) \leq \alpha \\
& \text{(b)}& \alpha \geq \alpha_{\mathrm{down}} \\
& \text{(c)}& \vec{0} \leq \vec{z} \leq \vec{z}_\mathrm{max} \quad,
\end{array} 
\label{eq:master} 
\end{equation}
\noindent where $l$ denotes the iteration of the master problem, $\alpha$ is a proxy for the subproblem costs, and the vector $\vec{\lambda}_\mathrm{s}$ is the weighted sum of the dual variables from the subproblems that are associated with the equalities in $\vec{z}$. The variable $J_\mathrm{sub}$ denotes the sum of the subproblem objective values. Constraint~(\ref{eq:master}a) represents the Benders cut at stage $l$ and constraints (\ref{eq:master}b,c) specify the bounds of the optimization variables.

\subsubsection{Sequential Subproblems}
\begin{figure}[!t]
	\centering
	\def\svgwidth{\columnwidth}
	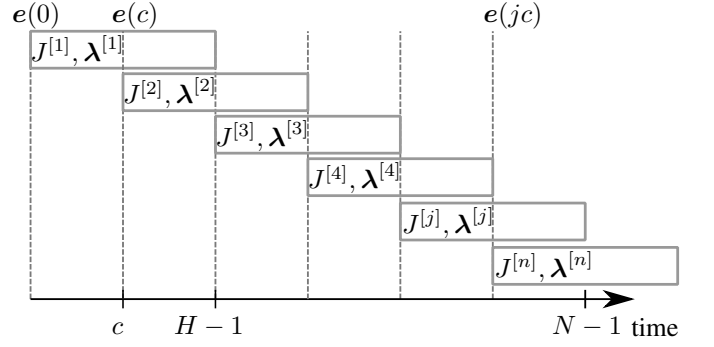
	\caption{Sequence diagram of subproblem decomposition for a closed loop \ac{MPC} strategy.}
	\label{fig:closedloop}
\end{figure}

To reflect an \ac{MPC} strategy we split the constraints in \eqref{eq:placesize}  with regard to the \ac{MPC} control horizon $H$ and to the partitions $\vec{X}=[\vec{x}^{[1]},\ldots,\vec{x}^{[n]}]^T$,$\vec{Y}=[\vec{y}^{[1]},\ldots,\vec{y}^{[n]}]^T$, and $\vec{D}=[\vec{d}^{[1]},\ldots,\vec{d}^{[n]}]^T$ by defining $\vec{A}^{[j]},\vec{b}^{[j]}$. The $j$-th coupled subproblem is defined as
\begin{equation}
\begin{array}{lll}
{J^*}^{[j]}(\vec{e}(jc),\vec{z})  = & \multicolumn{2}{l} {\underset{\vec{x}^{[j]},\vec{y}^{[j]},\vec{z},\vec{d}^{[j]} }{\min} \ T\left(\sum\limits_{k=0}^{H-1} \vec{1}^T \vec{y}^{[j]}(k) \right.} \\
	& &  \multicolumn{1}{c}{\left.+{\vec{c}^{[j]}_\mathrm{d}}^T \vec{d}^{[j]}(k)\right)}\\
& \text{s.t.}  \\
&\text{(a)}&  \vec{A}^{[j]}\left[ \begin{array}{c} \vec{x}^{[j]} \\ \vec{y}^{[j]} \\ \vec{z} \\ \vec{d}^{[j]} \end{array}\right]    \leq \vec{b}^{[j]}(\vec{e}(jc)) \\
& \text{(b)} & \vec{z} = \vec{z}^{(l)} : \vec{\lambda}^{[j]} \quad,
\label{eq:sub}
\end{array} 
\end{equation}
\noindent where $c$ denotes the update cycle of the subproblem recomputation. The coupling arises due to the fact that the actual \ac{SoE} vector from the previous optimization cycle needs to be transferred to the consecutive subproblem as initial state input denoted by $\vec{e}(jc)$. This fact does not allow us to solve the subproblems in a parallel fashion. The number of subproblems~$n$ depends on how often we rerun the optimization problems, which can be specified with the parameters $c$ and $N$. This is shown in Fig.~\ref{fig:closedloop}, in which the control actions are applied to the system for $c$ steps, before the next optimization problem is solved. This is also referred to as \ac{RHC} and can be regarded as a closed loop feedback applied to the system.	
To account for this strategy, we need to assign the individual contribution of each subproblem within $c$ steps in terms of the objective value and the dual variables $\vec{\lambda}^{[j]}$ that are associated with the equality constraints~(\ref{eq:sub}b). This can be achieved by calculating a weighted sum of $\vec{\lambda}^{[j]}$ with 
\begin{equation}
\vec{\lambda}_{\mathrm{s}} = \sum\limits_{j=1}^{n} \vec{\lambda}^{[j]}\frac{c}{H} \quad,
\label{eq:lambda_cL} 
\end{equation}
\noindent and determining the total objective value by
\begin{equation}
J_\mathrm{sub} = \sum\limits_{j=1}^n\sum\limits_{k=1}^{c} \ \vec{1}^T \vec{y}^{*[j]}(k) + {\vec{c}^{[j]}_\mathrm{d}}^T \vec{d}^{*[j]}(k) \quad,
\label{eq:jsub_cL} 
\end{equation}
\noindent where the total number of subproblems $n$ that have to be solved is
\begin{equation}
n = \frac{N}{c} \quad.
\label{eq:nsub} 
\end{equation}

\subsubsection{Algorithm}
With the aformentioned modifications we can solve the Benders decomposition in the same way as in \cite{Conejo2006}. Algorithm~\ref{benders_algo} specifies the needed steps.

\begin{algorithm}[t!]
	\begin{algorithmic}[1]
		\State $\vec{z} = \vec{0}, \ \alpha_\mathrm{down} = -100000, \ l=1$	
		\State  solve master problem \eqref{eq:master}  
		\Statex $[\vec{z}^{(l)},\alpha^{(l)}] = \arg\min J^{*} $ discarding (\ref{eq:master}a)
		\Do  
		\If {$(l> 1)$}
		\State calculate Benders cut (\ref{eq:master}a)
		\State solve master problem \eqref{eq:master} 
		\Statex  $[\setlength{\thickmuskip}{0mu} \vec{z}^{(l)},\alpha^{(l)}] = \arg\min J^{*}(J_\mathrm{sub}^{(l-1)},\vec{z}^{(l-1)},\vec{\lambda}_\mathrm{s}^{(l-1)}) $
		\EndIf
		\For{$j \ = 1 \ : \ n$}
			\State solve subproblem \eqref{eq:sub}  
			\Statex  $[\vec{x}^{[j]},\vec{z},\vec{d}^{[j]}] = \arg \min{J^*}^{[j]}(\vec{e}(jc),\vec{z})$ 
		\EndFor
		\State calculate $\vec{\lambda}_\mathrm{s}$ with \eqref{eq:lambda_cL} 
		\State calculate subproblem objective $J_\mathrm{{sub}}^{(l)}$ with \eqref{eq:jsub_cL}
		\State $Z_\mathrm{up}^{(l)} = J_\mathrm{{sub}}^{(l)} + \vec{c}_\mathrm{s}^T\vec{z}^{(l)}$
		\State $Z_\mathrm{down}^{(l)} = J_\mathrm{{sub}}^{(l)} + \alpha^{(l)}$
		\State $l = l + 1$
		\doWhile{$\left|\frac{Z_\mathrm{up}^{(l)}-Z_\mathrm{down}^{(l)}}{Z_{\mathrm{down}}^{(l)}}\right| > \epsilon$} \label{break}
	\end{algorithmic}
	\caption{Benders decomposition algorithm for optimal sizing and placement of distributed storage.}
	\label{benders_algo}
\end{algorithm}

Note that the only difference from \cite{Conejo2006} is that we have to solve the subproblems sequentially instead of in parallel.

\subsubsection{Problem Complexity}
We can also define a complexity bound for the decomposed problem (Alg.~\ref{benders_algo}). It can be approximated by adding up the runtimes of the subproblems~\eqref{eq:sub} as follows:
\begin{equation}
\mathcal{O}\left( l n H^3 (3 n_\mathrm{g} + 2 n_\mathrm{l} + n_\mathrm{b} + n_\mathrm{s})^3 \right) \quad  .
\end{equation}

\noindent This means, if we consider $n = N/H$ subproblems, we can achieve an acceleration by factor of $N^2/(l H^2)$ as compared to the problem~\eqref{eq:placesize}.

\section{Results}
\label{sec:results}
Here, we aim to define a realistic case study that assesses the economic value of different storage control strategies. 
\subsection{Test Case}
\begin{figure}[!t]
	\centering
	\includegraphics[width = \columnwidth]{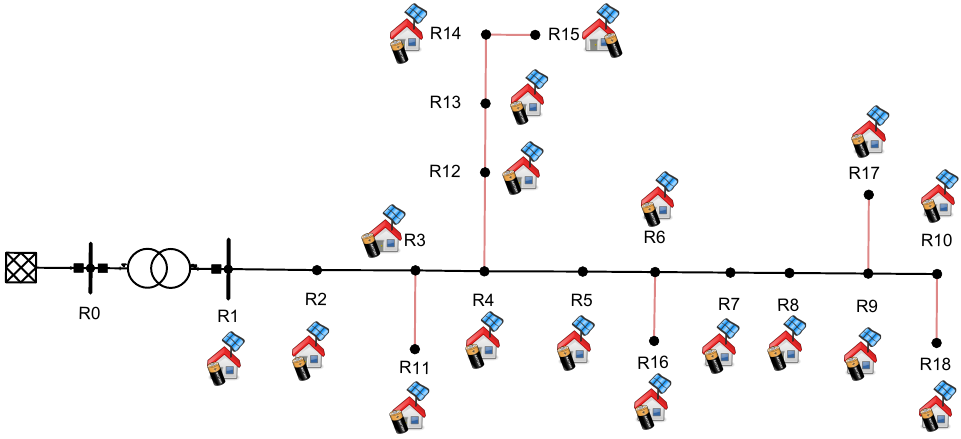}
	\caption{Local control area with group of prosumers (PV, battery systems and loads) populated on the CIGRE test grid from \cite{cigre}. The cables indicated with red lines have a higher resistance (2.05$\Omega$/km) than the black ones (0.405$\Omega$/km).}
	\label{fig:cigregrid}
\end{figure}

As depicted in Fig.~\ref{fig:cigregrid} we assume the \ac{LV} CIGRE benchmark grid \cite{cigre} as a realistic reference. The grid parameters are shown in Table~\ref{tab:linesetup}. As listed in Table~\ref{tab:sim_parameters} we configure the grid with a high PV penetration assuming that we can exploit the full roof top area of a single household.  We compute the optimal locations and sizes of the batteries for different \ac{MPC} controller strategies comprising different horizon lengths using the same input data for PV irradiation and load consumption. The controller objective is to minimize the generation costs at the feeder for a typical tariff scenario (day/night tariff \cite{EWZtarif} and feed-in tariff \cite{EWZfeedin}) in Switzerland indicated with the cost parameters $c_1^\mathrm{net},c_2^\mathrm{net}$ and $c_1^\mathrm{pv}$.  With this formulation we ensure that the group of prosumers gets rewarded for PV export and minimizes the consumption costs from the feeder by exploiting price differences of the tariff scheme and shifting energy from day to night through storage usage. In this way, the controller tries to utilize best the PV potential. We compute the placement and sizing problem for our simulation scenarios on the basis of just one full year to include the influence on seasonality, but also to save computation time. Therefore, $\vec{c}_\mathrm{s}$ represents the equivalent annual battery costs. In this regard we assume that the battery's calendar lifetime is at maximum 10 years.

\begin{table}[!t]
	\centering
	\caption{Line setup of the CIGRE test grid.}
	\begin{tabular}{ %
			@{}>{\rm\PBS\raggedright}p{1cm}@{}
			@{}>{\rm\PBS\raggedright}p{1cm}@{}
			@{}>{\rm\PBS\raggedright}p{2cm}@{}
			@{}>{\rm\PBS\raggedright}p{2cm}@{}
			@{}>{\rm\PBS\raggedright}p{1.1cm}@{}
			@{}>{\rm\PBS\raggedright}p{1.5cm}@{}}
		\hline
		Start node & End node & Resistance $R'$  [$\Omega$/km] & Reactance $X'_L$  [$\Omega$/km] &Length $l$ [m] & Max current $I_\mathrm{max}$ [A] \\ 
		\hline
		R1 & R2 & 0.405 & 0.205 & 35 & 398\\
		R2 & R3 & 0.405 & 0.205 & 35 & 398\\
		R3 & R4 & 0.405 & 0.205 & 35 & 398\\
		R4 & R5 & 0.405 & 0.205 & 35 & 398\\
		R5 & R6 & 0.405 & 0.205 & 35 & 398\\
		R6 & R7 & 0.405 & 0.205 & 35 & 398\\
		R7 & R8 & 0.405 & 0.205 & 35 & 398\\
		R8 & R9 & 0.405 & 0.205 & 35 & 398\\
		R9 & R10 & 0.405 & 0.205 & 35 & 398\\
		R3 & R11 & 2.05 & 0.212 & 35 & 158\\ 
		R4 & R12 & 2.05 & 0.212 & 30 & 158\\
		R12 & R13 & 2.05 & 0.212 & 35 & 158 \\
		R13 & R14 & 2.05 & 0.212 & 35 & 158\\
		R14 & R15 & 2.05 & 0.212 & 35 & 158\\
		R6 & R16 & 2.05 & 0.212 & 30 & 158\\
		R9 & R17 & 2.05 & 0.212 & 30 & 158\\
		R10 & R18 & 2.05 & 0.212 & 30 & 158\\
		\hline
	\end{tabular}
	\label{tab:linesetup}
\end{table}

\begin{table}[!t]
	\centering
	\caption{Simulation parameters.}
	\begin{tabular}{@{\hspace{2pt}}l@{\hspace{4pt}}l@{\hspace{2pt}}} \hline
		Storage units & 18 \\
		Storage power & $p_{\mathrm{gen}}^{\mathrm{s,max}}=$10kW, $q_{\mathrm{gen}}^{\mathrm{s,max}}=$ 10kVar \\
					  & rect. bounded \\  
		Storage efficiency & $\eta_\mathrm{dis}$ = 0.88, $\eta_\mathrm{ch} $= 0.88\\
		Degradation model & LiFePO4 convexified \\ 
						  &degradation map from \cite{Forman2012} \\
		Prediction horizon $H$ & 6h, 12h, 24h (1d), 168h (1w), 672h (1m)  \\
		Update cycle $c$ & 6h \\
		Sample time $T$ & 1h \\
		Feed-in tariff $c_1^\mathrm{net}: p_{\mathrm{gen}}^{\mathrm{net}}< 0$ &  50 \euro/MWh averaged from \cite{EWZfeedin}  \\
		Net power cost  $c_2^\mathrm{net}: p_{\mathrm{gen}}^{\mathrm{net}} \geq 0$ &  246 \euro/MWh 6:00-22:00 (Mon-Sat) \cite{EWZtarif}\\
									   &  131.5 \euro/MWh rest of time \\ 
		$\epsilon$ criterion & 0.01 \\					
		Battery cost $c_\mathrm{d}$ & 50-1000 \euro/kWh \\
		PV units & 18 \\
		PV power & $p_{\mathrm{gen}}^{\mathrm{pv,max}}$= 20 kW, \\
		& $q_{\mathrm{gen}}^{\mathrm{pv,max}}$ = 10kVar, rect. bounded  \\
		PV profiles & radiation profiles for the year 2015 \\ 
					& and the city of Zurich \\
		PV gen cost $c_1^\mathrm{pv}$ &  0 \euro/MWh \\
		Storage gen cost $c_1^\mathrm{s}$ & 0 \euro/MWh \\
		Total PV production & 465 MWh \\
		Total load consumption & 61.5 MWh \\
		Simulation horizon $N$ & 8760   (1 year)\\
		Households & 18 @ 4kWp  generated \\
				   & load profiles from \cite{bucher} \\
		Grid & European \ac{LV} network \cite{cigre} \\
		Voltage limits & $v_\mathrm{max} = 1.1$,$v_\mathrm{min} = 0.9$ \\
		Thermal limits & according to \cite{cigre} \\
		\hline
	\end{tabular}
	\label{tab:sim_parameters}
\end{table}

\begin{figure*}[t]
	\centering
	\def\svgwidth{\textwidth}
	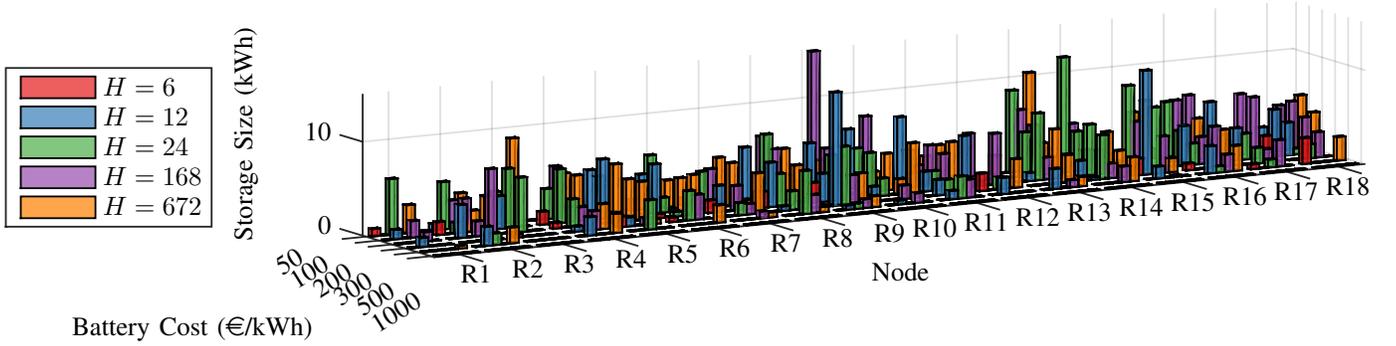
	\caption{Optimal placement and sizing results of the installed distributed battery storage as a function of the battery investment cost and control horizon $H$. The height of the bars represent the storage size, while the battery locations correspond to the nodes referenced in Fig.~\ref{fig:cigregrid}. The control horizon $H$ also corresponds to hours.}  
	\label{fig:placement}
\end{figure*}

\subsection{Heuristic Controller}
To compare our enhanced predictive storage control strategy, we modify a standard heuristic control strategy that is described in \cite{Weniger2014,Marra2014}. In particular, we consider the storage control strategy from \cite{Weniger2014} that has a fixed feed-in limitation. The authors of \cite{Marra2014} refer to this mode as a conventional storage strategy. This rule-based controller does not include any forecast of the PV production and is therefore a non-predictive controller. It stores surplus PV power during the day and curtails PV power when the batteries are full and the grid limit is exceeded. In contrast to \cite{Weniger2014} and with the aim to utilize more PV power, we force the batteries to empty in the morning in cases when the available energy content was not consumed by the household over night. In addition, we determine dynamically the grid limits by running an \ac{AC-OPF}. The \ac{AC-OPF} framework allows us to also consider optimal \ac{RPC} and \ac{APC}. 

Unfortunately, we cannot formulate an optimal placement and sizing problem using the proposed Benders decomposition technique, since this strategy does not provide any dual variables to reduce the space of feasible solutions by Benders cuts. However, to compare the strategies, we run the heuristic strategy with the optimal storage configuration that is obtained by the \ac{MPC} strategy.      

\subsection{Convergence and Computation Time}
\begin{figure}[t]
	\centering
	\includegraphics[width = \columnwidth]{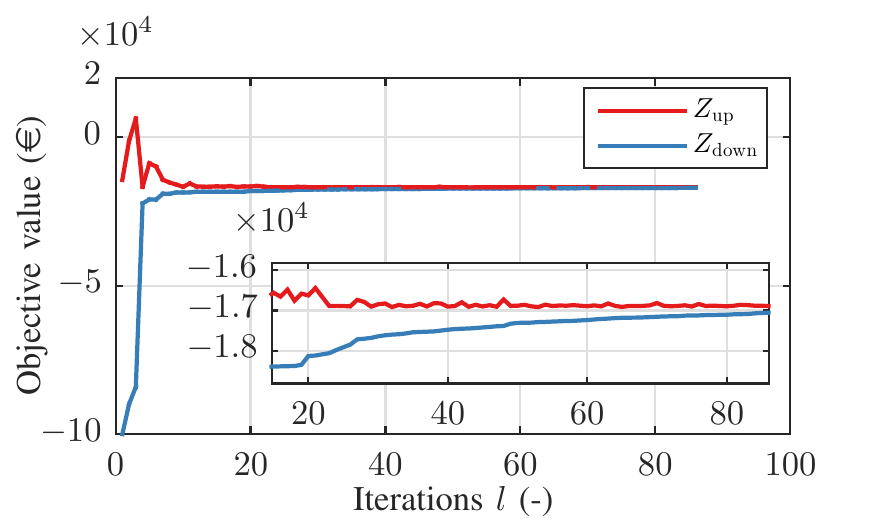}
	\caption{Typical convergence rate of the proposed Benders decomposition algorithm.}
	\label{fig:iterations}
\end{figure}

Figure~\ref{fig:iterations} shows the typical convergence rate of our proposed Benders decomposition approach for one simulation scenario. A simulation scenario is defined as a full-year simulation ($N$=8760) at a given control horizon $H$ and a fixed battery cost level. One can observe that around 80 iterations are needed to reach the $\epsilon$-criterion specified in Alg.~\ref{benders_algo}. The computation time ranges from 24 hours ($H=24$) to 3 days ($H=672$) to sequentially solve one simulation scenario with the CPLEX \ac{LP} solver \cite{cplex}. Instead of solving all 30 simulation scenarios consecutively, we run multiple simulation scenarios in a parallel fashion on multi-core processors to save computation time.  

\subsection{Sizing and Placement}
First, we run the sizing and placement problem without the degradation model in the \ac{MPC} strategy to further save computation time. Figure~\ref{fig:placement} shows the results of the optimal \ac{DBS} distribution as a function of different control horizons and battery costs. The highest line loading we observe is at line R1-R2. The largest storage sizes are placed at the nodes R12-R15 and R9 to support the line R1-R2. The placement decisions for R12-R15 are associated with lines that have a higher resistance. Also the resistance from the feeder to the node R9 is higher due to a longer cable length. This means that with this configuration we can reduce the network losses most effectively and therefore utilize more PV power. 

\begin{figure}[t]
	\centering
	\includegraphics[width = \columnwidth]{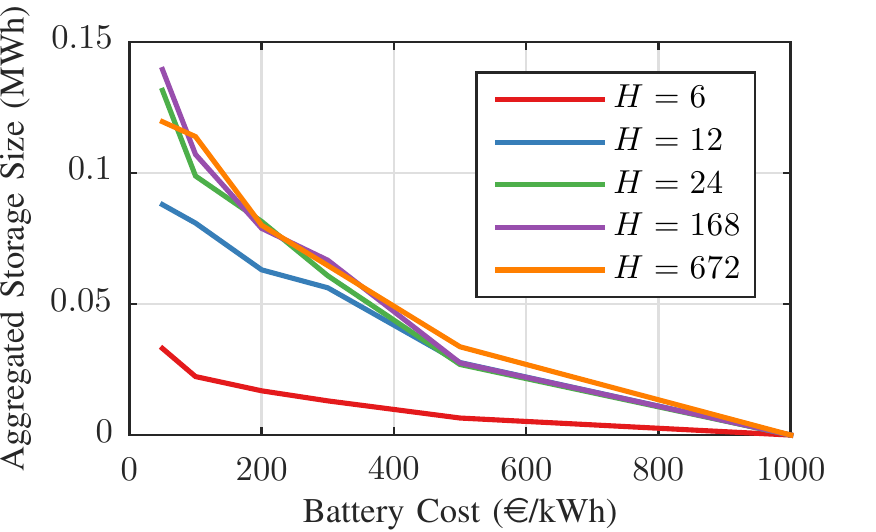}
	\caption{Aggregated storage size as a function of the battery cost for different control horizons.}
	\label{fig:aggsize}
\end{figure}

Figure~\ref{fig:aggsize} shows the aggregated placed storage size as a function of the battery cost. One can see that when decreasing the horizon length, less storage capacity is placed. In addition, at a cost level of 1000\euro/kWh it is not viable any longer to place storage in the grid. It is noteworthy that the storage size is almost the same, whether operating the storage at daily ($H=24$), weekly ($H=168$) or monthly ($H=672$) control horizons.   

\begin{figure}[!t]
	\centering
	\includegraphics[width = \columnwidth]{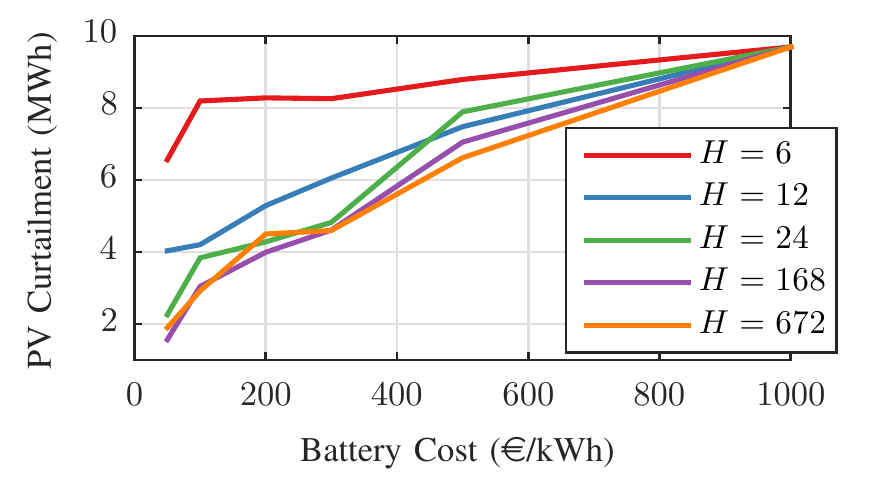}
	\caption{Total PV curtailment as a function of the battery cost for different control horizons.}
	\label{fig:aggcurtail}
\end{figure}

Fig.~\ref{fig:aggcurtail} shows the PV curtailment as a function of the battery cost for the different control horizons. One can observe that the curtailment levels are higher, when using sub-daily horizons. By using longer horizons ($H=24,168,672$) the PV curtailment is reduced by about half, since the aggregated storage size is higher for these controller strategies (see Fig.~\ref{fig:aggsize}). 

\begin{figure}[!t]
	\centering
	\includegraphics[width = \columnwidth]{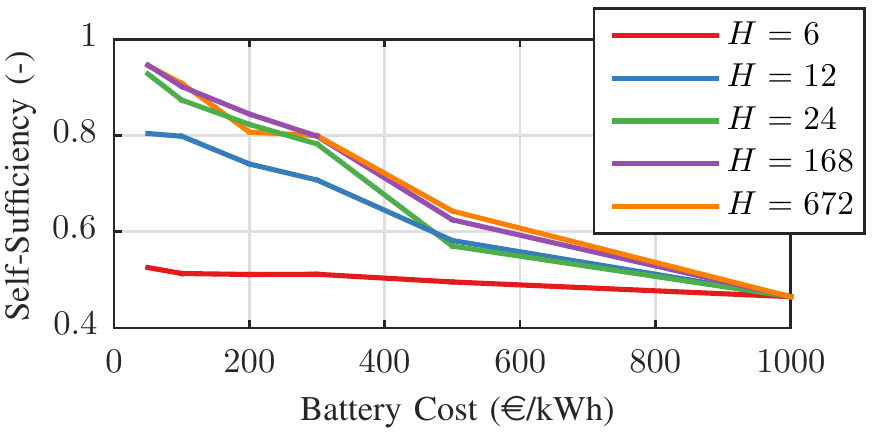}
	\caption{Self-sufficiency of the control area as a function of the battery cost for different control horizons.}
	\label{fig:selfSuff}
\end{figure}

However, despite of the same storage size, the degree of self-sufficiency of the local control area for the horizon strategies ($H=168,672$) is in almost all cases higher (see Fig.\ref{fig:selfSuff}). It can be anticipated that the economic benefit of shifting energy over weekly or monthly intervals to cover the loads is not significant for the considered battery cost levels. This means that the driving factor for sizing is the mitigation of PV curtailment on daily patterns, such that the optimal size is determined by finding the best compromise between minimizing PV curtailment and battery investment. This is also the reason why we observe a saturation in storage size for multi-day horizons, since the storage size is already large enough to increase the self-sufficiency. The degree of self-sufficiency describes to which extent the local control area is independent of the grid. It is defined as 
\begin{equation}
\text{self-sufficiency} = \frac{E_\mathrm{ld}-E_\mathrm{net}^\mathrm{im}}{E_\mathrm{ld}} \quad,
\end{equation} 
\noindent where $E_\mathrm{ld}$ is the yearly energy consumption of all loads and $E_\mathrm{net}^\mathrm{im}$ is the yearly imported energy from the feeder. A factor of one would mean that the control area is off-the-grid.      

\subsection{Economic Value of Horizon Length}
\begin{figure}[!t]
	\centering
	\includegraphics[width = \columnwidth]{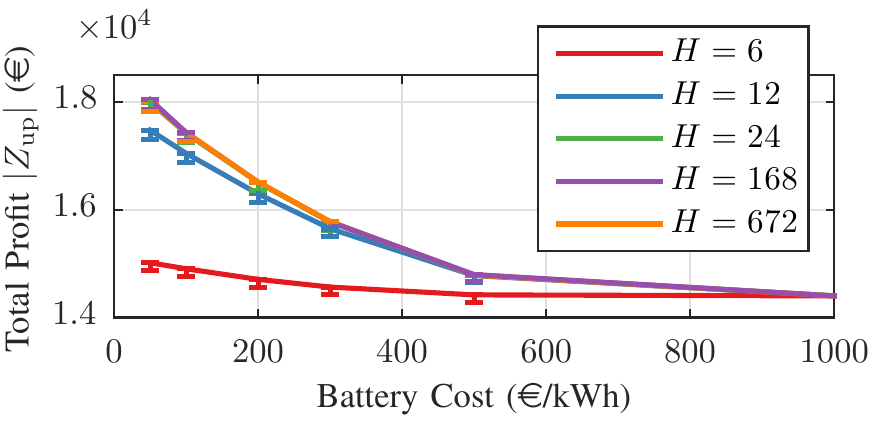}
	\caption{Yearly investment profit as a function of the battery cost for different control horizons.}
	\label{fig:objectiveHorizon}
\end{figure}

Next, we aim to compare the impact on the investment profit for different control horizon lengths. Figure~\ref{fig:objectiveHorizon} shows the investment profit corresponding to the objective value of \eqref{eq:master} as a function of the battery cost for different control horizons. One can observe that sub-daily horizon strategies are less profitable than daily, weekly or monthly horizon strategies. Another interesting result is that multi-day horizons perform similarly, which means that longer control horizons than one day do not further improve the profit.           

\subsection{Economic Assessment}
\begin{figure}[!t]
	\centering
	\includegraphics[width = \columnwidth]{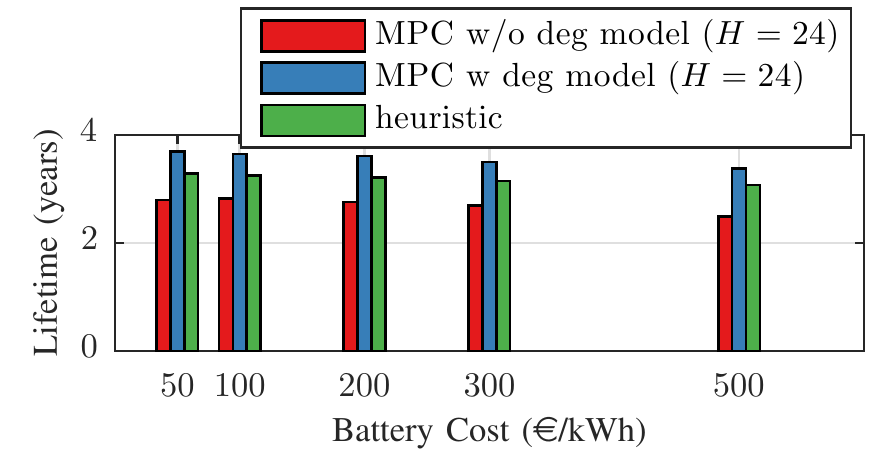}
	\caption{Battery lifetime as a function of the battery cost for different storage control strategies for an EoL criterion 0.8. Note that no battery capacity is installed for the battery cost of 1000\euro/kWh.}
	\label{fig:lifetime}
\end{figure}

\begin{figure}[!t]
	\centering
	\includegraphics[width = \columnwidth]{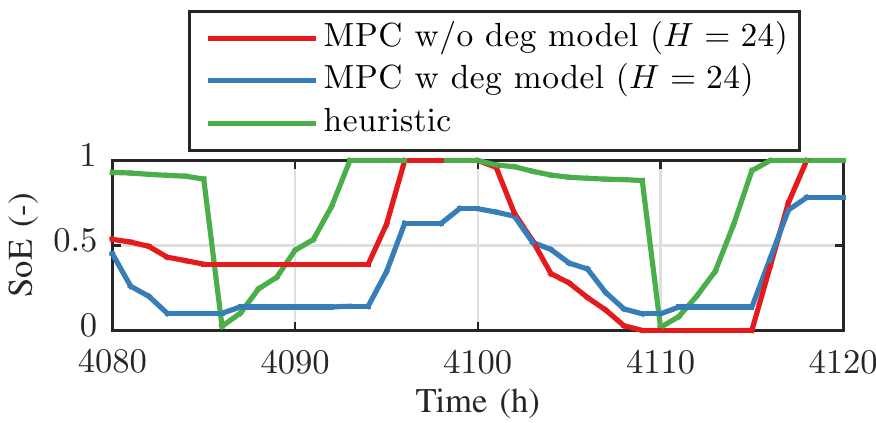}
	\caption{Resulting state of energy (SoE) trajectories from different storage control strategies.}
	\label{fig:controlCompare}
\end{figure}

For the economic assessment, we compare the \ac{MPC} strategies with and without degradation model for a 24 hour horizon with the heuristic storage control strategy. By using the degradation map \eqref{eq:hull}, we first compute the battery lifetime in $m$ years for the different strategies according to the \ac{EoL} criterion of 0.8. Here, the \ac{EoL} defines the aggregated remaining capacity in the given control area after $m$ years. This allows us to assume that the expected revenue streams over the years are approximately the same. Figure~\ref{fig:lifetime} shows the battery lifetimes for the type of different storage control strategies. 
It can be inferred that the storage control strategy has a great impact on battery lifetime. This can be explained by Fig.~\ref{fig:controlCompare} that shows the resulting evolution of the \ac{SoE} trajectories from the different controllers. While the \ac{MPC} with degradation model avoids \ac{SoE} regimes that are associated with high battery wear, the other ones idle the batteries at low \ac{SoE} regimes or use the full capacity potential. 
The variable $-J_\mathrm{sub}$ can be regarded as the revenue stream for one year. To account only for the storage investment, we need to define the revenue difference considering an investment with (wS) and without storage (w/oS). The \ac{NPV} of the investment is
\begin{equation}
\text{NPV} =  -\vec{c}_\mathrm{d}^T\vec{z} + \sum\limits_{k=1}^m\frac{-J_\mathrm{sub}^\text{wS}+J_\mathrm{sub}^\text{w/oS}}{(1+\text{IRR})^k}
\label{eq:IRR} \quad.
\end{equation}
To obtain a viable investment the \ac{NPV} has to be greater than zero. Since the \ac{IRR} is a direct measure for the \ac{ROI}, we solve \eqref{eq:IRR} for the \ac{IRR} by setting the \ac{NPV} to zero.

\begin{figure}[!t]
	\centering
	\includegraphics[width = \columnwidth]{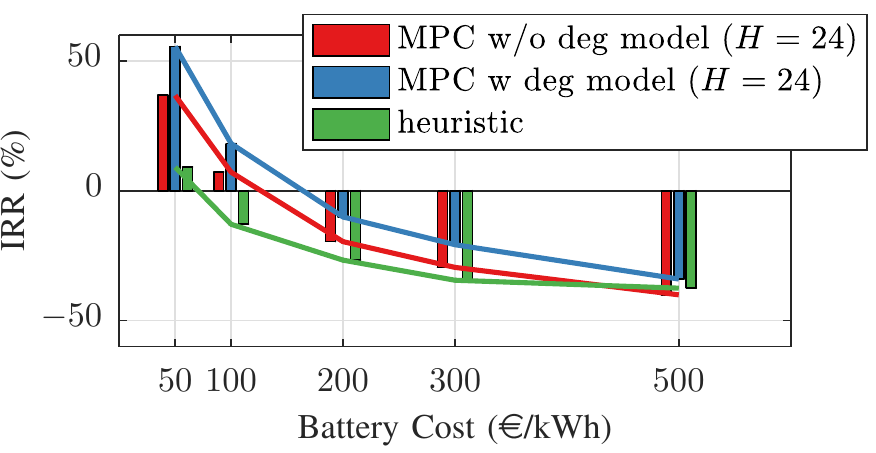}
	\caption{Internal Rate of Return (IRR) as a function of the battery cost for different storage control strategies.}
	\label{fig:IRR}
\end{figure}

Although the battery lifetime is longer for the heuristic controller as compared to the \ac{MPC} strategy without degradation model, the \acp{IRR} for the \ac{MPC} strategies are superior, which are shown in Fig.~\ref{fig:IRR}. When using \ac{MPC} strategies the group of prosumers gets viable results below battery cost levels of $\approx$175\euro/kWh (w deg model) and $\approx$125\euro/kWh (w/o deg model), while the heuristic strategy only achieves a profit below $\approx$60\euro/kWh. This is due to the fact that the \ac{MPC} strategies can generate more value by using forecast information and therefore better utilize the batteries.       

\section{Conclusion}
\label{sec:conclusion}
This paper presents a novel Benders decomposition method that considers \ac{MPC} strategies in a planning and operation problem. We split the sizing and placement problem into a master planning problem and sequentially-solvable subproblems reflecting a predictive storage control strategy. The storage control strategy is formulated as an \ac{MPC} strategy that optimally schedules distributed battery storage to maximize PV self-sufficiency and PV utilization while considering grid constraints and minimizing battery degradation. The grid constraints are incorporated as a multi-period \ac{OPF} problem using an existing linearized version of the \ac{OPF}. Due to the linear property of the placement and sizing problem, it can be decomposed by using Benders decomposition. 

From the case study it can be concluded that \ac{MPC} strategies are in general more profitable than heuristic controller strategies and are viable for battery costs below 175\euro/kWh. The horizon length has a great impact on profitability. Control horizons that are shorter than 24 hours limit the revenue potential, while operating storage on a daily base is as good as on a weekly or monthly base in terms of the overall profitability. Nevertheless, higher horizon lengths increase the degree of self-sufficiency of the control area. The main conclusion on the optimal sizing and placement of \ac{DBS} is that the best scheme is achieved when overloaded network elements are supported and network losses are reduced. As a further finding, by using a battery degradation model within our \ac{MPC} controllers, we can extend the battery lifetime and hence further increase the total profitability of the group of prosumers.

Future work relates to further analyze the impact of different PV installations and network topologies on the optimal placement and sizing of DBS. In this regard, it could also be studied whether centralized or distributed battery storage configurations are more preferable.


\bibliographystyle{IEEEtran}
\bibliography{literature}
\end{document}